\documentclass[aps,pra,twocolumn,showpacs,preprintnumbers,amsmath,amssymb,footinbib,longbibliography,superscriptaddress,10pt]{revtex4-2}
\usepackage[utf8]{inputenc}
\usepackage[english]{babel}
\usepackage{lmodern}
\usepackage{empheq}
\usepackage{mathtools} 
\usepackage[toc,page]{appendix}
\usepackage{graphicx}
\usepackage{physics}
\usepackage{mleftright}
\usepackage{hyperref}
\usepackage[dvipsnames]{xcolor}
\usepackage[section]{placeins}

\usepackage{hyperref}
\definecolor{darkblue}{rgb}{0.031,0.282, 0.49} 
\definecolor{link_red}{rgb}{0.686, 0.016, 0} 
\hypersetup{
    colorlinks      = true,
    linkcolor       = link_red,
    menucolor       = black,
    citecolor       = link_red,
    urlcolor        = link_red,
}

\newcommand{\figref}[1]{\mbox{Fig.~\ref{#1}}}

\newcommand{\secref}[1]{\mbox{Section~\ref{#1}}}

\newcommand{\appref}[1]{\mbox{Appendix~\ref{#1}}}
\renewcommand{\eqref}[1]{\mbox{Eq.~(\ref{#1})}}

\newcommand{\figpanel}[2]{Fig.~\hyperref[#1]{\ref*{#1}(#2)}}
\newcommand{\figpanels}[3]{Fig.~\hyperref[#1]{\ref*{#1}(#2)-(#3)}}
\newcommand{\figpanelNoPrefix}[2]{\hyperref[#1]{\ref*{#1}(#2)}}
\newcommand{\figlink}[2]{\hyperref[#1]{\color[rgb]{0.031,0.282, 0.49}#2}}

\newcommand{\round}[1]{\mleft(#1\mright)}
\newcommand{\hard}[1]{\mleft[#1\mright]}
\newcommand{\curly}[1]{\mleft\{#1\mright\}}
\renewcommand{\exp}[1]{{\rm exp}\hard{#1}}
\renewcommand{\Tr}[1]{{\rm Tr}\mleft[ #1 \mright]}

\renewcommand{\min}[1]{{\rm min}\mleft\{ #1 \mright\}}
\renewcommand{\max}[1]{{\rm max}\mleft\{ #1 \mright\}}

\newcommand{\R}{{\rm R}}
\renewcommand{\L}{{\hat{L}}}
\renewcommand{\H}{{\hat{H}}}
\newcommand{\rhoh}{{\hat{\rho}}}
\renewcommand{\S}{\hat{S}}
\newcommand{\Sx}{\hat{S}^x}
\newcommand{\Sy}{\hat{S}^y}
\newcommand{\Sz}{\hat{S}^z}

\newcommand{\Pois}{{\rm Pois}}

\newcommand{\Erl}{{\rm Erlang}}
\newcommand{\Pt}{ {\rm P}_t }
\newcommand{\Pm}{ {\rm P}_m }

\newcommand{\mt}{{\tilde m}}
\newcommand{\jump}[2]{#1 #2 #1^\dagger}
\newcommand{\rel}[1]{\Re{#1}}
\newcommand{\iml}[1]{\Im{#1}}

\newcommand{\xt}{\tilde{x}}

\newcommand{\cg}[1]{\textcolor{Orange}{#1}}

\begin{document}

\title{Conditions for implementing projective measurements through continuous monitoring
}

\author{Therese~Karmstrand}
\email[Email:]{karmstrand@gmail.com}
\affiliation{Center for Quantum Computing, RIKEN, Wako-shi, Saitama 351-0198, Japan}
\author{Franco~Nori}
\affiliation{Center for Quantum Computing, RIKEN, Wako-shi, Saitama 351-0198, Japan}
\affiliation{Physics Department, University of Michigan, Ann Arbor, MI 48109-1040, USA}
\author{Clemens~Gneiting}
\affiliation{Center for Quantum Computing, RIKEN, Wako-shi, Saitama 351-0198, Japan}
%

\begin{abstract}
Projective measurements are foundational for quantum physics in general and quantum information tasks in particular. However, their direct implementation is often not warranted. Here, we investigate under what conditions continuous monitoring that manifests in quantum jump trajectories realises projective measurements over time. Considering finite-dimensional quantum systems and single, diagonalisable jump operators, we show if and how the detected quantum jump statistics force, and allow inferring, the convergence of individual quantum trajectories towards eigenstates of an observable in the long-time limit. We identify a necessary non-degeneracy condition that is related to the presence of a strong symmetry with non-degenerate symmetry sectors. We derive analytical expressions for the rate of convergence and the time-error relationship in finite-time measurements. Our results provide a transparent framework for understanding the emergence of projective measurements from continuous monitoring, with direct implications for the optimisation of quantum measurement protocols in experiments.
\end{abstract}
\maketitle

\section{Introduction}
Projective measurements have played a foundational role in quantum theory since their conceptualisation~\cite{neumann_mathematische_1996,schrdinger_gegenwrtige_1935, gleason_measures_1957}, linking elusive wave functions to tangible experimental outcomes through the Born rule. While modern quantum measurement theory has matured into the more general framework of positive operator-valued measures (POVMs)~\cite{wiseman_quantum_2010,jacobs_quantum_2014,davies_operational_1970,srinivas_photon_1981,ozawa_quantum_1984,ozawa_uncertainty_2004,ashhab_weak_2009,ashhab_information_2009,ashhab_information_charge_qubit_2009,dressel_certainty_2014}, projective measurements retain a distinguished status because of their repeatability and their ability to prepare well-defined post-measurement states according to Lüders' rule~\cite{wiseman_quantum_2010}. They therefore remain a central resource both in theory and experiments~\cite{steffen_deterministic_2013,steffinlongo_projective_2022,khandelwal_simulating_2025,yamamoto_measurement-induced_2026}.

Yet, despite their ubiquitous appearance in quantum theory, ideal projective measurements are often difficult, or even impossible, to implement directly in the laboratory. Across many experimental platforms, information is instead acquired indirectly and continuously~\cite{vijay_stabilizing_2012, murch_observing_2013, weber_mapping_2014, tan_prediction_2015, minev_catch_2019} through interaction between the system and an ancilla, or environment. Two standard measurement techniques are photon counting and homodyne detection, both of which produce continuous stochastic measurement records whose infinitesimal time steps are described by non-projective POVMs. While sufficiently long and efficient continuous measurements are routinely employed to realise high-fidelity readout~\cite{chen_transmon_2023,myerson_high-fidelity_2008,guerlin_progressive_2007}, for example in circuit QED via dispersive qubit measurement~\cite{filipp_two-qubit_2009,krantz_single-shot_2016,walter_rapid_2017,krantz_quantum_2019,wang_ideal_2019,spring_fast_2025}, it remains important to clarify under what conditions the underlying stochastic dynamics implement an ideal projective measurement satisfying both the Born and Lüders' rules.

Continuous measurements are naturally described by quantum trajectories, in which the conditional quantum state evolves according to the observed measurement record~\cite{srinivas_photon_1981,tian_quantum_1992,carmichael_open_1993,wiseman_quantum_2010,jacobs_quantum_2014}. Discarding the measurement results recovers a corresponding Lindblad master equation.
While the same master equation can be associated with different measurement schemes and, hence, a one-to-one correspondence is lost, some conditions necessary for the emulation of projective measurements can already be understood on the level of the ensemble evolution. For instance, ensemble dynamics that drive the system into a unique steady state are evidently incompatible with the Born rule. To satisfy the Born rule, clearly a pure dephasing process is necessary. But even if the ensemble dynamics describe a pure dephasing process, whether or not the quantum trajectories converge to an eigenstate of an observable, as imposed by Lüders' rule, ultimately depends on the details of the quantum jump operators that characterise the measurement, and thus goes beyond the ensemble description. In this work, we inquire about the additional conditions on the jump operators and the associated measurement to emulate an ideal projective measurement.

In the case of quantum-{\it jump} trajectories, the conditional state evolves through a stochastic sequence of discrete detector clicks~\cite{scully_quantum_1997,carmichael_statistical_2002,wiseman_quantum_2010,agarwal_quantum_2012}. In the extreme case of only a single monitored jump operator, the temporal succession of its click record constitutes the only information available about the quantum state. Whether the measurement asymptotically realises an ideal projective measurement must therefore be encoded entirely in the counting statistics of the detected jumps. This raises an obvious but decisive question: how can we warrant that this information is sufficient for identifying a unique eigenstate?

To address this question, we investigate a general, not necessarily Hermitian, jump operator that is diagonal in the eigenbasis of the Hamiltonian. This guarantees a pure dephasing process when discarding the measurement records, as required by the Born rule. Exploiting the full counting statistics~\cite{menczel_full_2026} of the detection record, we derive necessary and sufficient conditions for convergence, and obtain convergence rates and confidence estimates at finite times. Moreover, adding constant shifts to the jump operator allows us to interpolate towards the limit of diffusive measurements, where discrete quantum jumps resolve into a continuous Wiener process. Finally, we show how the first-passage-time distribution of the quantum jumps, which describes the distribution of arrival times at a fixed jump count $m$ and thus complements the full counting statistics,  gives rise to analogous convergence conditions and rates.

The convergence of quantum trajectories to emulate projective measurements was addressed in previous works. For instance, in a pioneering work by Wiseman and Wilburn~\cite{wiseman_quantum_1993}, asymptotic convergence towards the eigenstates was shown for the case where the jump operator coincides with a field quadrature of a field mode.   Observables with continuous spectra were also studied in the strong-measurement limit in Ref.~\cite{bauer_monitoring_2018}.
In Ref.~\cite{bauer_convergence_2011}, it was shown that repeated, discrete-time quantum non-demolition measurements converge to an eigenstate of the measured observable, and was shortly generalised to continuous time in Ref.~\cite{bauer_repeated_2013}.
The case of general diagonal jump operators in finite-dimensional Hilbert spaces was treated in Ref.~\cite{benoist_large_2014}, where similar convergence conditions and rates were found. In contrast to Ref.~\cite{benoist_large_2014}, we base our reasoning on the full counting statistics (and the first-passage time distribution) of a single jump operator, which allows us to develop an intuitive understanding of the convergence process. Moreover, while discrete jump and diffusive Wiener processes are discussed in parallel in Ref.~\cite{benoist_large_2014}, we treat the latter as a limiting case of the former, thereby further elucidating their relationship in the context of trajectory convergence.

The asymptotic behaviour of quantum trajectories has also been studied in the context of ergodicity. We here refer to ergodicity in the conventional thermodynamic sense, where an individual quantum trajectory is ergodic if its time-average in the long-time limit equals the ensemble average over all possible trajectories, that is, the solution of the corresponding master equation. It has been shown for finite-dimensional systems~\cite{kummerer_pathwise_2004} that individual quantum trajectories are ergodic if the master equation admits a unique steady-state solution.

On the other hand, when multiple solutions exist, the trajectory's time-average converges to \textit{one} of these states, selected at random. The latter implies that single trajectories may be \textit{non-ergodic}, as the ensemble-average quantum state in this case would be a mixture of the multiple steady states, determined by the initial conditions. This phenomenon has been coined ``dissipative freezing''~\cite{sanchez_munoz_symmetries_2019} and has recently been discussed as a general property of systems with a strong symmetry, except for two exceptions~\cite{tindall_generality_2023}. Our work is consistent with, and provides new perspectives on, these results. Specifically, we find that the asymptotic realisation of a projective measurement implies non-ergodic quantum trajectories if the state is initially in a non-trivial superposition of different eigenstates of the jump operator. Moreover, we find that the necessary non-degeneracy conditions can be associated with the existence of a strong symmetry with non-degenerate symmetry sectors.

\section{Quantum Jump Trajectories}\label{Sec:quantum-trajectories}
Often quantum jump trajectories are motivated by, and derived from, the Gorini–Kossakowski–Sudarshan–Lindblad (GKSL) or Lindblad master equation ~\cite{carmichael_statistical_2002, gardiner_quantum_2010}. In this approach, the solution of the Lindblad equation is expressed as a Dyson series, which is then interpreted as an ensemble average over quantum trajectories. In the situation considered here, monitoring through weak continuous measurement is the starting point, and quantum trajectories emerge directly as a consequence of the conditioning on the measurement record~\cite{carmichael_statistical_2002, wiseman_quantum_2010, wiseman_quantum_1993}.

If the nature of the continuous measurement is such that it induces piecewise deterministic dynamics, interspersed by stochastically occurring quantum jumps (sometimes called a ``point'' or ``jump'' process), the associated continuous binary measurement can be characterised by the measurement operators \cite{wiseman_quantum_2010}
\begin{subequations} \label{eq:weak-measurement}
	\begin{align}
		\hat{M}_1(dt) &= \L\sqrt{\gamma dt} , \\
		\hat{M}_0(dt) &= \hat{\mathbb{I}} - \left(\frac{i}{\hbar} \H + \frac{\gamma}{2} \L^\dagger \L \right) dt ,
	\end{align}
\end{subequations}
where the jump operator $\L$ captures the measurement back-action induced by the quantum jumps, which manifest as detector ``clicks''. Here, we consider the case of a single jump operator, but the generalisation to a set of jump operators is straightforward. Note that the inherent system dynamics described by the Hamiltonian $\H$ is, for convenience, subsumed in the effect of the ``null outcome''.

Performing the measurement (\ref{eq:weak-measurement}) and discarding the measurement outcome results in the state update
\begin{align}
	\rhoh(t+dt) = \hat{M}_0(dt) \rhoh(t) \hat{M}_0^\dagger(dt)+\hat{M}_1(dt) \rhoh(t) \hat{M}_1^\dagger(dt) 
\end{align}
which, to first order in $dt$, recovers the standard Lindblad equation
\begin{align} \label{eq:Lindblad_equation}
	\frac{d }{dt} \rhoh(t) = -\frac{i}{\hbar}\comm{\H}{\rhoh} + \gamma \round{\jump{\L}{\rhoh} - \ \frac{1}{2} \acomm{\L^\dagger \L}{\rhoh}} . 
\end{align}
If the measurement record is kept, on the other hand, the state is conditioned on the occurrences of the quantum jumps. After a finite measurement time $t$, the conditioned states, which are referred to as quantum trajectories, are then given by
\begin{subequations} \label{eq:quantum_trajectory}
\begin{align}
	\label{eq:conditional-state}
	\rhoh_c(t) = \frac{\rho_c(t)}{\Tr{\rho_c(t)}} ,
\end{align}
where 
\begin{align}
	\label{eq:unnormalised-conditional-state}
	\rho_c(t) = \mathcal{S}(t-t_m)\mathcal{J}\mathcal{S}(t_m-t_{m-1})...\mathcal{J}\mathcal{S}(t_1)\rhoh(0)
\end{align}
is an unnormalised trajectory and 
\begin{align}
	\label{eq:probability-density-m}
	p(t_1,t_2,\cdots,t_m;[0,t]) = \Tr{\rho_c(t)}
\end{align}
\end{subequations}
is the exclusive probability density for detecting $m$ quantum jumps, each within the disjoint infinitesimal time intervals $[t_1,t_1+dt_1],[t_2,t_2+dt_2],\cdots,[t_m,t_m+dt_m]$, in the measurement time $[0,t]$. A quantum jump is described by the superoperator $\mathcal{J}$ with the action
\begin{align}
    \label{eq:jump}
    \mathcal{J}\rhoh = \gamma \L \rhoh \L^\dagger ,
\end{align}
which corresponds to a ``click'' of the detector. The dominant ``null'' outcome (conditioned on the absence of quantum jumps), on the other hand, gives rise to intervals of deterministic evolution in between the quantum jumps, described by the superoperator~\cite{srinivas_photon_1981,wiseman_quantum_1993}
\begin{align}
    \label{eq:smooth}
    \mathcal{S}(t)\rhoh = \exp{\hat{Y}t} \rhoh \, \exp{\hat{Y}^\dagger t} , 
\end{align}
with $\hat{Y}\coloneq -\tfrac{i}{\hbar}\H-\tfrac{\gamma}{2}\L^\dagger\L$.

In terms of the quantum trajectories (\ref{eq:quantum_trajectory}), the solution to the Lindblad equation (\ref{eq:Lindblad_equation}) can now be written as the ensemble average~\cite{wiseman_quantum_1993, carmichael_statistical_2002}
\begin{widetext}
\begin{equation}
    \label{eq:dyson}
    \rhoh(t) = 
        \sum_{m=0}^\infty \int_{0}^{t}\dd t_m \int_{0}^{t_{m}}\dd t_{m-1}  ... \\
        \int_{0}^{t_2}\dd t_1 \mathcal{S}(t-t_m)\mathcal{J}\mathcal{S}(t_m-t_{m-1})...\mathcal{J}\mathcal{S}(t_1)\rhoh(0) ,
\end{equation}
which is also obtained by solving (\ref{eq:Lindblad_equation}) through a formal Dyson series. Let us remark that detecting the quantum jumps also enables an alternative averaging protocol, where quantum trajectories are averaged at a fixed jump count $m$ (instead of at a fixed time $t$), while the time required to arrive at $m$ varies from trajectory to trajectory~\cite{gneiting_jumptime_2021,gneiting_unraveling_2022}. The corresponding ensemble-averaged state
\begin{equation}
	\label{Eq:jumptime-averaged_state}
	\rhoh_m =  \int_{0}^{\infty}\dd t_m \int_{0}^{t_{m}}\dd t_{m-1}  ... \\
	\int_{0}^{t_2}\dd t_1 \mathcal{J}\mathcal{S}(t_m-t_{m-1})...\mathcal{J}\mathcal{S}(t_1)\rhoh(0)
\end{equation}
\end{widetext}
is then not governed by the Lindblad equation (\ref{eq:Lindblad_equation}), but by the jump-time evolution equation, which describes a discrete quantum map from $m$ to $m+1$ \cite{gneiting_jumptime_2021}. Depending on whether the readout time $t$ or the jump count $m$ is kept fixed, either $m$ or $t$ becomes a random variable, giving rise to the full counting statistics and the first-passage-time statistics, respectively \cite{menczel_full_2026}.

In this work, we inquire under which circumstances the weak measurement (\ref{eq:weak-measurement}) implements an ideal projective measurement. That is, we inquire if and how the stochastic quantum jump process, over time, projects the system state onto a single eigenstate of an observable. 
To this end,  we first take the fixed-$t$ perspective and subsequently demonstrate that the fixed-$m$ conditioning gives rise to analogous results.

\section{Conditions for implementing  an ideal projective measurement}
\label{sec:implementing-a-projective-measurement}

Projective measurements refer to { observables} $\hat{\Lambda}$, which are Hermitian operators that, expressed in terms of their spectrum, read
\begin{equation}
    \hat{\Lambda}= \sum_\lambda \lambda \hat{\Pi}_\lambda ,
\end{equation}
where the real eigenvalues $\lambda$ stand for the measurement outcomes, and the $\hat{\Pi}_\lambda$ are projectors on the corresponding eigenspaces. As projection operators they satisfy the defining property $\hat{\Pi}_\lambda^2 = \hat{\Pi}_\lambda$. If all eigenvalues $\lambda$ are non-degenerate, we have rank-1 projectors $\hat{\Pi}_\lambda = |\psi_\lambda \rangle \langle \psi_\lambda|$, in which case the measurement is sometimes called {ideal projective measurement} or {von Neumann measurement}. In the general case, $\hat{\Pi}_\lambda = \sum_{j=1}^{N_\lambda} |\psi_{\lambda, j} \rangle \langle \psi_{\lambda, j}|$, where the degeneracy $N_\lambda$ of the eigenvalue $\lambda$ determines the rank of the associated projector.

Projective measurements are characterised by Born's outcome probability rule and L{\"u}ders' state update rule. Born's rule states that the probability ${\rm Pr}(\lambda)$ to detect outcome $\lambda$ is given by
\begin{align} \label{Eq:Born_rule}
    {\rm Pr}(\lambda) = {\rm Tr}[\hat{\Pi}_\lambda \hat{\rho}] , \hspace{7mm} \text{(Born's rule)}
\end{align}
where $\hat{\rho}$ is the state that enters the measurement. L{\"u}ders' state update rule (projection postulate) then states that, conditioned on the measurement outcome $\lambda$, the system's post-measurement state is
\begin{align} \label{Eq:Lueders_rule}
    \hat{\rho}_\lambda' = \frac{\hat{\Pi}_\lambda \hat{\rho} \hat{\Pi}_\lambda}{{\rm Pr}(\lambda)} . \hspace{12mm} \text{(L{\"u}ders' rule)}
\end{align}
In particular, if the outcome $\lambda$ is nondegenerate, then $\hat{\rho}_\lambda' = |\psi_\lambda \rangle \langle \psi_\lambda|$, that is, the measurement prepares a pure state that is completely determined by the measurement outcome. Moreover, it follows immediately from (\ref{Eq:Born_rule}) and (\ref{Eq:Lueders_rule}) that a subsequent measurement of the same observable delivers the same measurement outcome (repeatability), if the time between the measurements is sufficiently short or if $[\hat{H}, \hat{\Pi}_\lambda] = 0$. Non-projective POVMs, including the weak measurement (\ref{eq:weak-measurement}), in general lack these properties. In the language of POVMs, projective measurements are distinguished by measurement operators $\hat{M}_\lambda$ and effects $\hat{M}_\lambda^\dagger\hat{M}_\lambda$ that coincide with projection operators $\hat{\Pi}_\lambda$: $\hat{M}_\lambda = \hat{M}_\lambda^\dagger\hat{M}_\lambda = \hat{\Pi}_\lambda$.

The repeatability condition indicates that projective measurements belong to the broader class of {nondestructive} or {non-demolition} measurements. Nondestructive measurements are POVMs which, when the measurement outcomes are discarded, describe a pure dephasing process; that is, they preserve the populations in a selected basis, in accordance with Born's rule. Continuous measurements such as (\ref{eq:weak-measurement}), to implement a projective measurement in the long-time limit, must thus necessarily be nondestructive. Below we will argue that this is warranted if the jump operator $\hat{L}$ is diagonalisable and commutes with the Hamiltonian. Photon detections, on the other hand, which deplete the state towards the vacuum, represent ``destructive'' measurements. But even if a continuous measurement is nondestructive, it is not guaranteed that individual quantum trajectories converge towards a state or subspace of the selected basis. A counterexample, a pure dephasing process, where trajectories do not converge, is discussed in \secref{sec:qubit-sz-measurement}. One of the goals of this article is to establish the additional conditions on the jump operator $\hat{L}$ that warrant such convergence.

Formally, we can identify a non-destructive measurement as a measurement of an observable that is a constant of motion in the Heisenberg picture. By inspection of the measurement operators 
in \eqref{eq:weak-measurement}, we may identify the effect $\hat{M}_1^\dagger\hat{M}_1(dt) \propto  \L^\dagger \L$ as a constant of motion, and therefore identify the continuous measurement process as a non-destructive measurement of $\L^\dagger \L$, if
\begin{itemize}
    \item[(i)] there exist a complete, orthonormal basis $\curly{\ket{i}}$ in which the operator $\L$ is diagonal, and 
    \item[(ii)] the Hamiltonian $\H$ is simultaneously diagonal in the same basis, or equivalently $\comm{\H}{\L}=0$. 
\end{itemize}
The first condition refers to the existence of a set of orthogonal pointer states that are invariant under the measurement $\hat{M}_1$, and the second ensures that these states are also invariant under the null measurement $\hat{M}_0$. 
In fact, for finite-dimensional Hilbert spaces, it can be rigorously proven that the stochastic master equation corresponding to measurement (\ref{eq:weak-measurement}) describes a non-destructive measurement process in the basis $\curly{\ket{i}}$ if and only if both $\H$ and $\L$ are diagonal in $\curly{\ket{i}}$, see Ref.~\cite{benoist_large_2014} (theorem 3).
We additionally note that conditions (i) and (ii) define both the measurement operator $\hat{M}_1 \propto \L$ and the effect $\hat{M}_1^\dagger\hat{M}_1(dt) \propto  \L^\dagger \L$ as strong symmetries. However, as we argue in \secref{sec:strong-symmetry}, only the latter is relevant for guiding the asymptotic behaviour of individual trajectories.

Phrased differently, the conditions (i) and (ii) constrain the measurement process (\ref{eq:weak-measurement}) to a pure dephasing process, and it is clear that the process preserves the initial-state statistics associated to the observable $\L^\dagger \L$. Hence, a hypothetical ideal projective measurement on the conditional state in the basis $\curly{\ket{i}}$ would produce the state $\ket{i}$ with a probability $\Tr{\hat{\Pi}_i \rhoh_c(t)}$ ascribed by the Born rule, exactly as if the measurement had been performed on the initial state itself. In our case, however, the effect $\hat{M}_1^\dagger\hat{M}_1$ represents a weighted sum over \textit{all} projection operators onto the eigenstates $\curly{\ket{i}}$, which distinguishes this measurement process from a direct implementation where each eigenspace projector coincides with a separate measurement operator. Due to this difference, conditions (i) and (ii) alone are not sufficient to warrant that the measurement process (\ref{eq:weak-measurement}) will converge the conditional state towards a single eigenstate of $\L^\dagger \L$.

In the following, assuming that conditions (i) and (ii) are satisfied for the measurement (\ref{eq:weak-measurement}), we identify a necessary third, non-degeneracy condition, which guarantees that each trajectory selectively converges to a single eigenstate of $\L^\dagger \L$, hence implementing an ideal projective measurement.  
We demonstrate this for an arbitrary $d$-dimensional Hilbert space with a discrete set of basis states $\curly{\ket{i}}_{i=1}^d$. 
Our starting point are the quantum-jump trajectories (\ref{eq:quantum_trajectory}), and our analysis follows an approach similar to the non-destructive quadrature measurement studied in Ref.~\cite{wiseman_quantum_1993}, where our work can be seen as a generalisation to a generic discrete-variable non-destructive measurement. The key objects for our demonstration are the probability distribution $\Pt(m)$ and the conditional distributions $\Pt(i | m )$, which are derived below.

\subsection{Quantum jump statistics}
Let $\curly{\ket{i}}$ be a complete set of orthonormal basis states on a $d$-dimensional Hilbert space and assume $[\H,\L]=0$ such that $\L\ket{i}=\sum_{i=1}^dl_i\ket{i}$ ($l_i \in \mathbb{C}$), and  $\H\ket{i}=\sum_{i=1}^dh_i\ket{i}$.  In the basis $\curly{\ket{i}}$,  the actions of $\mathcal{J}$ and $\mathcal{S}(t)$ [Eqs.~(\ref{eq:jump}) and (\ref{eq:smooth})] evaluate as
\begin{subequations}
\begin{align}
    \label{eq:action-jump}
    & \mathcal{J} \ketbra{i} = \gamma \abs{l_i}^2 \ketbra{i}, \\
    & \mathcal{S}(t) \ketbra{i} = \exp{-\gamma t \abs{l_i}^2} \ketbra{i}.
    \label{eq:action-smooth}
\end{align}
\end{subequations}
The condition $\big[\H,\L\big]=0$ additionally ensures that $\mathcal{J}$ and $\mathcal{S}(t)$ commute. It follows that the conditional state depends only on the total number of jumps $m$, and not on their times of occurrence. Formally, this allows us to write the unnormalised trajectory in the simpler form 
\begin{equation}
    \rho_c(t) = \mathcal{S}(t)\mathcal{J}^m\rhoh(0). 
\end{equation}
Next, by integrating the exclusive probability density corresponding to this state over all possible successive detection times of $m$ jumps in the time $t$, we obtain an expression for the probability distribution of the random variable $m$,
\begin{align}
    \label{eq:Ptm-general-expression}
    \Pt(m) &= \int_0^{t} dt_m \cdots \int_0^{t_2}dt_1 \Tr{\mathcal{S}(t)\mathcal{J}^m\rhoh(0)}.
\end{align}
Evaluating the trace using the actions in Eqs.~(\ref{eq:action-jump}) and (\ref{eq:action-smooth}), and performing the time integrals, yields the expression
\begin{subequations}     \label{eq:Ptm}
\begin{align}
        \Pt(m) &=  \sum_{i=1}^d p_i \frac{(\gamma t  \abs{l_i}^{2})^m}{m!} \exp{-\gamma t \abs{l_i}^2},\\
        &=\sum_{i=1}^d p_i \Pois(\lambda_i;m),        \label{eq:Ptm-poisson}
\end{align}
\end{subequations}
where the index $t$ is used to denote the total measurement time.

As illustrated by \eqref{eq:Ptm-poisson}, the distribution of $m$ is a sum of Poisson distributions with mean values $\lambda_i \coloneq \gamma t \abs{l_i}^2$ that are weighted by the probabilities $p_i = \mel{i}{\rhoh(0)}{i}$ in the initial state. Each of these Poissonian modes 
 is directly associated with a different eigenvalue of the observable $\L^\dagger\L=\sum_{i=1}^d \abs{l_i}^2 \ketbra{i}$, which suggests that the continuous measurement may become interpretable as a measurement of the observable $\L^\dagger\L$ once all modes are statistically distinguishable. To see this, note that a hypothetical ideal projective measurement on the conditional state in the basis $\curly{\ket{i}}$ would turn the outcome $\abs{l_i}^2$ itself into a random variable.
 The distribution $\Pt(m)$ is then recovered as the marginal distribution of the joint distribution
\begin{equation}\label{eq:Ptlm-joint}    
    \Pt(i,m)=p_i \frac{(\gamma t)^m}{m!}  \abs{l_i}^{2m} \exp{-\gamma t \abs{l_i}^2}
\end{equation}
for the random variables $m$ and $\abs{l_i}^2$, where we have used the simpler label $i$ for the outcome $\abs{l_i}^2$.
Similarly, the probability of obtaining the result $\abs{l_i}^2$ is given by the complementary marginal
\begin{equation}
    \label{eq:Pti}
    \Pt(i) = \sum_{m=0}^\infty  \Pt(i,m) = p_i = \mel{i}{\rhoh(0)}{i},
\end{equation} 
recovering the Born rule under a hypothetical ideal projective measurement, in accordance with $\L^\dagger \L$ being a constant of motion.
From Eqs.\,(\ref{eq:Ptm}) and (\ref{eq:Ptlm-joint}) one also readily derives the conditional probability distribution for the outcome $\abs{l_i}^2$ given the jump count $m$ as
\begin{equation}
\label{eq:Ptcond}
    \Pt(i |m) = \frac{p_i \abs{l_i}^{2m} \exp{-\gamma t \abs{l_i}^2} }{\sum_{i=1}^d p_i \abs{l_i}^{2m} \exp{-\gamma t \abs{l_i}^2}}.
\end{equation}

\begin{figure*}
    \centering
    \includegraphics[width=\linewidth]{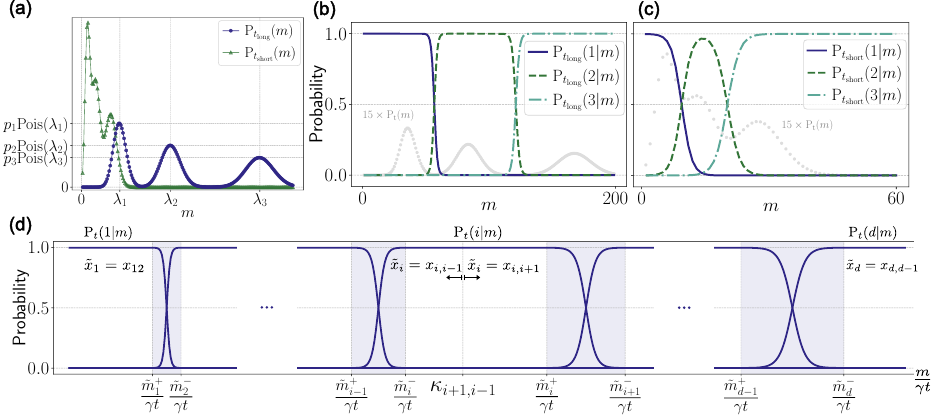}
    \caption{[(a),(b),(c)] Sample probability distributions for a 3-level system ($d=3$): (a) distribution of jumps $\Pt(m)$, and conditional probability distributions $\Pt(i|m)$ for (b) $t=t_{\rm long}$, and (c)  $t=t_{\rm short}$. The parameters used to produce this figure were: $\abs{l_1}^2=0.3$, $\abs{l_2}^2=0.7$, $\abs{l_3}^2=1.4$, $p1=p2=p3=\tfrac{1}{3}$, $ \gamma t_{\rm long} = 120 $,  $\gamma t_{\rm short} =20 $.
    (d) Illustration of $\sum_i \Pt(i|m)$ for an arbitrary $d$-level system, highlighting the intervals in which we show that the cumulative probabilities $\Pt(\mt_{i}^+ \leq m \leq \mt_{i+1}^-)$ vanish in the long time limit. The figure also indicates $\xt_i=\max{x_{ij}}_{j\neq i}$, which governs the functional behaviour of each of the $\Pt(i |m)$, in the regions of interest. }
    \label{fig:Proof-illustration}
\end{figure*}

These distributions are key to demonstrating the realisation of a projective measurement. Specifically, we note that a projection onto the state $\ket{i}$ is implied when $\Pt(i | m)=1$ for some measurement time $t$, and all other $\Pt(j \neq i | m)=0$.
The expression in \eqref{eq:Ptcond} reveals an intricate interplay between $m$ and $t$ that governs the conditional probability. To gain a better intuition for the functional dependence of this interplay, we factor out the numerator and define the variables
\begin{equation}
    \label{eq:xij}
    x_{ij}= \gamma t \round{\abs{l_i}^2\hspace{-1pt} -\abs{l_j}^2}\hspace{-0.5pt} - m \ln{\frac{\abs{l_i}^2}{\abs{l_j}^2}} \hspace{-0.5pt} - \ln{\frac{p_i}{p_j}},
\end{equation}
which allows us to write the conditional probabilities as
\begin{equation}
\label{eq:Ptcond_xij}
    \Pt(i |m) = \frac{1}{1+\displaystyle \sum_{j\neq i} \exp{x_{ij}}}.
\end{equation}
Depending on the relationship between $m$ and $t$, and for sufficiently long measurement time $t$, there will be distinct intervals of $m$ where only one of the $\exp{x_{ij}}$ is dominant, making the conditional probabilities behave as logistic functions, having s-shaped transitions between 0 and 1, and vice versa when one dominant $\exp{x_{ij}}$ gives way for another. This functional behaviour is illustrated for a three-level system with distinct eigenvalues $\abs{l_i}^2$ in \figpanel{fig:Proof-illustration}{b}. In this figure, we additionally plot a magnification of the probability distribution $\Pt(m)$, which associates each of the distinct regions where $\Pt(i | m)=1$ to the distinguishability of the corresponding Poissonian modes identified in \eqref{eq:Ptm-poisson}. In contrast, for insufficiently long measurement time $t$, these modes greatly overlap and prevent a single $\exp{x_{ij}}$ from dominating the conditional probabilities.  Clearly, non-degeneracy among the eigenvalues $\abs{l_i}^2$ is also crucial for distinguishing the different modes over time. In the trivial situation where all eigenvalues are equal, $\abs{l_i}^2=\abs{l}^2,\, \forall i$, the probability distribution $\Pt(m)$ reduces to a single Poissonian, and the conditional probabilities are simply given by the initial conditions, $\Pt(i |m)=p_i$. 
Indeed, we may formulate a necessary non-degeneracy condition for the realisation of a projective measurement through the continuous measurement (\ref{eq:weak-measurement}) as follows.

\subsubsection{Non-degeneracy condition} \label{sec:non-degeneracy-condition}
\noindent
\textit{Let $\curly{\ket{i}}$ be a finite set of orthonormal eigenstates of $\L$, $\L\ket{i}=l_i \ket{i}$, and assume $\big[ \H, \L \big]=0$.  
If the initial state is not already in an eigenstate of $\L$, a quantum jump trajectory realised by the measurement (\ref{eq:weak-measurement}) may converge to the eigenstate $\ket{i}$ iff the initial overlap with this state is non-zero, $p_i>0$, and the eigenvalues of $\L^\dagger \L$ satisfy}
\begin{align}
\label{eq:non-degeneracy-condition}
     \abs{l_i}^2 \neq \abs{l_j}^2, \quad \forall j\neq i.
\end{align}
\noindent
This condition is necessary to enable the asymptotic projection onto the state $\ket{i}$, but does not demonstrate that convergence to a single state will occur with certainty.
Showing this is the goal of the next section.

\subsection{Asymptotic convergence to a single eigenstate} \label{sec:asymptotic-convergence}
Here, we demonstrate that all trajectories converge to a single, but randomly chosen, state $|i \rangle$ in the long-time limit, given that the non-degeneracy condition above holds for all $\ket{i}$. That is, we now assume that all eigenvalues of $\L^\dagger \L$ are distinct. Without loss of generality, we label the states such that $\abs{l_1}^2<\abs{l_2}^2 < \cdots < \abs{l_d}^2$. 
It is also clear from \eqref{eq:Ptcond} that the conditional state will not evolve in time if the initial state is already in an eigenstate $\ket{i}$ and consequently $p_i =1$, and that the conditional dynamics cannot reach the state $\ket{i}$ if there is no overlap with that state initially. We therefore only consider $p_i\in(0,1), \; \forall i$.

\subsubsection{Convergence condition}
A single trajectory can be said to converge to a single eigenstate if
\begin{equation}
\label{eq:convergence}
    \lim_{t\to \infty} \frac{1}{d-1}\sum_{i=1}^{d-1}\sum_{j=i+1}^d \abs{\Pt( i |m)-\Pt(j |m)}= 1, 
\end{equation}
as the pairwise difference takes on the values
\begin{equation} \label{eq:pairwise-difference}
    \abs{\Pt\round{ i |m}-\Pt\round{ j |m}} = 
        \begin{cases}
            1 & \text{if} \; \ket{\psi}_t=\ket{i \,{\rm or}\,j}, \\
            0 & \text{if} \; \ket{\psi}_t=\ket{k \neq i,j}, \\
            \in\mleft(0,1\mright) & \text{otherwise}. \\
        \end{cases}
\end{equation}
Here $\ket{\psi}_t$ denotes the pure conditional state given the measurement count $m$ and measurement time $t$. \\

To demonstrate convergence, we need to show that $\lim_{t\to \infty}\Pt(i|m)=1$ or 0, $\forall i $.
As discussed in the previous section, the existence of this limit is directly linked to the statistical distinguishability of the Poissonian modes constituting $\Pt(m)$. In particular, we want to show that the cumulative probability for obtaining a jump count $m$ in the intervals where the different $\Pt(i|m)$ cross, vanishes in the long-time limit. See the shaded areas in \figpanel{fig:Proof-illustration}{d}.
To achieve this, we first note that 
\begin{align}
	\label{eq:Pcond-max}
	\Pt(i | m) \geq \frac{1}{1+(d-1)\exp{\xt_i} } 
\end{align}
where $\xt_i = \max{x_{ij_1}, x_{ij_2}, ..., x_{ij_{d-1}}}$ and the $x_{ij}$ are defined as in \eqref{eq:xij}. Next, we introduce an arbitrarily small error $\varepsilon$ and require that 
\begin{align}
	\label{eq:Pcond-bound}
	\Pt(i | m) \geq 1-\varepsilon.
\end{align}
By substituting $\xt_i = \xt_\varepsilon$ in \eqref{eq:Pcond-max} and solving for equality with $1-\varepsilon$ gives the threshold
\begin{equation}
	\label{eq:xep}
	\xt_\varepsilon = \ln{\frac{\varepsilon}{(d-1)(1-\varepsilon)}},
\end{equation}
which achieves the bound in \eqref{eq:Pcond-bound}. For long measuring times $t$, it is also possible to show that 
\begin{align}
\label{eq:xmax}
	\xt_i = 
	\begin{cases}
		&x_{i,i-1}, \quad  \text{for}\; \; \kappa_{i-2,i-1} <\frac{m}{\gamma t } < \kappa_{i-1,i+1}, \\
		&x_{i,i+1}, \quad \text{for}\; \; \kappa_{i-1,i+1} < \frac{m}{\gamma t } < \kappa_{i+1,i+2}, 
	\end{cases}
\end{align}
which covers the interval where $\Pt( i | m) \approx 1 $. See \appref{sec:app-xmax} for details.
The parameter $\kappa_{ij}$ is defined as
\begin{equation}
\label{eq:kij}
	\kappa_{ij} = \frac{ \abs{l_i}^2 -\abs{l_j}^2 }{\ln{ \frac{\abs{l_i}^2}{\abs{l_j}^2} }}.
\end{equation}
Then, by solving $x_{i,i\pm1}=\xt_\varepsilon$, we obtain the bounding jump counts 
\begin{equation}
	\label{eq:mipm}
	\mt_i^\pm = \frac{\gamma t \round{\abs{l_i}^2- \abs{l_{i\pm1}}^2 } -\xt_\varepsilon - \ln{\frac{p_i}{ p_{i\pm1}}} }{ \ln{ \frac{ \abs{l_i}^2 }{ \abs{l_{i\pm1}}^2 } } },
\end{equation}
which define the interval $\mt_i^- < m < \mt_i^+$ within which the bound (\ref{eq:Pcond-bound}) is satisfied. For the edge modes corresponding to the smallest and largest eigenvalues, the lower and upper bounds are $\mt_1^-=0$ and $\mt_d^+ \to {\infty}$, respectively.

Equipped with this characterisation, we proceed and evaluate the cumulative probability of finding $m$ outside these intervals. That is, the intervals where the conditional probabilities cross. For a $d$-level system with $d$ distinct eigenvalues, there are ($d-1$) crossings in total, which we label with the index $i\in(1,d-1)$. Employing \eqref{eq:Ptm-poisson}, we find that
\begin{align} \label{eq:cumulative-bound}
		\Pt(\mt_i^+\hspace{-0.07cm} \leq \hspace{-0.03cm}m \hspace{-0.03cm}\leq\hspace{-0.03cm} \mt_{i+1}^- )
		& \leq \hspace{-0.3cm}\sum_{k: \lambda_k < \mt_i^+} \hspace{-0.2cm}p_k \Pois( \lambda_k; m \geq \mt_{i}^+ ) \nonumber \\
		& \; + \hspace{-0.45cm} \sum_{k: \lambda_k > \mt_{i+1}^-} \hspace{-0.35cm} p_k \Pois( \lambda_k; m \leq \mt_{i+1}^- )  
\end{align}
The two cumulative tail probabilities of the Poisson distributions, found on the right-hand side of this inequality, are in turn bounded by a Chernoff bound~\cite{mitzenmacher_probability_2005} according to
\begin{subequations}
\begin{align}
    &\Pois\round{\lambda_k; m \geq  \mt_{i}^+ } \leq \frac{({\rm e} {\lambda_k})^{ \mt_{i}^+ }\exp{-\lambda_k}}{(\mt_{i}^+)^{\mt_{i}^+}}, \; \lambda_k < \mt_{i}^+,\\
    &\Pois\round{\lambda_k; \hspace{-0.035cm} m  \hspace{-0.035cm}\leq  \hspace{-0.035cm}\mt_{i+1}^-  } \hspace{-0.06cm}\leq\hspace{-0.06cm} \frac{({\rm e} {\lambda_k})^{{\mt_{i+1}^-} }\exp{-\lambda_k}}{ ({\mt_{i+1}^-})^{\mt_{i+1}^-}}, \; \lambda_k \hspace{-0.05cm}>\hspace{-0.04cm}\mt_{i+1}^-.
\end{align}
\label{eq:tail-bounds}
\end{subequations}
In \appref{app:vanishing-tails}, we show that $\lambda_k < \mt_i^+$, $\forall \lambda_k \leq \lambda_i$ and $\lambda_k > \mt_{i+1}^-$, $\forall \lambda_k \geq \lambda_{i+1}$ for long measurement times $ t$, and that the right-hand sides of \eqref{eq:tail-bounds} vanish for all intervals $i$ when we take the limit $t \to \infty$ . Thus, we find that
\begin{align}
    \label{eq:lim-Ptm-cumulative}
    \lim_{\varepsilon \to 0} \lim_{t\to \infty} \sum_{i=1}^{d-1} \Pt(\mt_i^+ \leq m \leq \mt_{i+1}^- ) = 0,
\end{align}
which was the goal of this section. 

The above result demonstrates that any realisation of the measurement process (\ref{eq:weak-measurement}) will, in the long-time limit, generate jump counts $m$ that fall within one of the intervals $\mt_i^- \leq m \leq \mt_i^+$, where $\Pt(i|m)\geq1-\varepsilon$, with an arbitrarily small error $\varepsilon$, with unit probability. It follows that the convergence condition in \eqref{eq:convergence} is satisfied, which proves that all trajectories asymptotically converge to a single eigenstate of $\L^\dagger\L$. By accomplishing this, we have shown that the continuous measurement process (\ref{eq:weak-measurement}) asymptotically implements a projective measurement of the observable $\L^\dagger\L$ when $[\H,\L]=0$ and the non-degeneracy condition (\ref{eq:non-degeneracy-condition}) is fulfilled $\forall i$. That is, a hypothetical projective measurement of the observable $\L^\dagger\L$ on the conditional post-measurement state, at times sufficiently large to warrant convergence towards an eigenvalue $|l_i|^2$, would deliver the state $|i\rangle$, in accordance with Lüders' rule (\ref{Eq:Lueders_rule}). Moreover, since $\L^\dagger\L$ is a constant of motion, the probability of obtaining the result $\abs{l_i}^2$ is given by the Born rule as if the measurement was performed on the initial state. See \eqref{eq:Pti}.

\begin{figure}
    \centering
    \includegraphics[width=.85\linewidth]{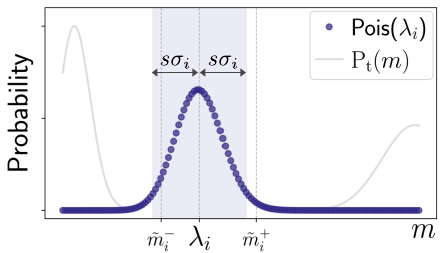}
    \caption{ Illustration of the condition for the minimum measurement time for convergence to $\ket{i}$ in terms of $s$ standard deviations. In this example, $s=2$ and $\varepsilon=10^{-3}$, and one sees that the lower tail is the limiting factor due to the closer distance to $\Pois(\lambda_{i-1};m)$. }  
    \label{fig:confidence-level}
\end{figure}

\subsection{Convergence rate and the time-error relationship} \label{sec:time-error-relationship}
A realistic implementation of a projective measurement through continuous monitoring is necessarily performed within a finite measurement time. Therefore, it is natural to ask how long it takes to arrive at an outcome within a small but non-zero error $\varepsilon$ such that $\Pt(i|m)$ satisfies the bound in \eqref{eq:Pcond-bound}.

Above, we required that $\xt_i=\xt_\varepsilon$, which gave the bounding jump counts $\mt_i^\pm$ presented in \eqref{eq:mipm}. Moreover, to show that each trajectory converges to a single eigenstate, we also required that $\lambda_i < \mt_i^+$ and $\lambda_{i+1} > \mt_{i+1}^-$, for $i\in[1,d-1]$, which can be extended to $\mt_i^- < \lambda_i < \mt_i^+$, for $i\in[1,d]$ by recognising that the lower and upper bounds for $i=1,d$ are automatically satisfied as $\mt_1^-=0$ and $\mt_d^+\to \infty$. 

In our proof, we assumed an arbitrarily small error $\varepsilon$ and took the limit $t \to \infty$. Now, we instead inquire the minimum measurement time to attain the result $\ket{i}$ with $\varepsilon$ small but finite. This precision is statistically achieved on average when $\mt_i^- < \lambda_i < \mt_i^+$. Accordingly, we find the convergence time $t_i(\varepsilon)$ by solving the two inequalities $\mt_i^- < \lambda_i$ and $\lambda_i < \mt_i^+$ for $t $ and maximising over the two results: 
\begin{align}
\label{eq:convergence-time}
	\gamma t_i(\varepsilon) \geq \max{ \frac{\abs{\xt_\varepsilon}-\ln{\frac{p_i}{p_{i\pm1}}}}{ \abs{l_{i\pm1}}^2 - \abs{l_i}^2 +  \abs{l_i}^2\ln{\frac{\abs{l_i}^2}{\abs{l_{i\pm1}}^2}}  }, 0},
\end{align}
where $0$ is the maximum only when $-\ln{p_i/p_{i\pm1}}>\abs{\xt_\varepsilon}$.
Here, we also see that unequal initial probabilities act as an offset that can reduce or prolong the convergence time. The corresponding convergence rate is identified as
\begin{align}
\label{eq:convergence-rate}
	\R_i = \min{ \abs{l_{i\pm1}}^2 - \abs{l_i}^2 +  \abs{l_i}^2\ln{\frac{\abs{l_i}^2}{\abs{l_{i\pm1}}^2}} }.
\end{align}
While we obtain this rate via a conceptually different method, it coincides with the convergence rate found for jump trajectories in Ref.~\cite{benoist_large_2014}. However, in contrast to~\cite{benoist_large_2014}, where the rate compares the eigenvalue $\abs{l_i}^2$ to any of the other eigenvalues and does not specify which rate is the slowest, our method allows us to identify the distinguishability to the nearest neighbouring modes as the limiting factor to the convergence, wherefore we identify the convergence rate as the slowest rate among those two.

Comparing the convergence time in \eqref{eq:convergence-time} with numerical simulations shows, on average, a good agreement between measurement time and error. 
The confidence level depends on the choice of parameters, but is generally found to be around $40-50\%$, since the conditions $\mt_i^- < \lambda_i$, and $\lambda_i < \mt_i^+$  imply that at least the upper or lower half of $\Pois(\lambda_i;m)$ fall within the interval of jump counts $m$ that would give $\Pt(i|m) \geq 1-\varepsilon$, whichever represents the slowest time scale. To increase the confidence level, one can require that a larger fraction of $\Pois(\lambda_{i}; m)$ satisfy the respective condition. Hence, a natural way to quantify the confidence level is through numbers $s$ of standard deviations, $s \sigma_{i}$,  where $\sigma_i = \sqrt{\lambda_i}$ for the Poisson distribution. See \figref{fig:confidence-level}. The corresponding conditions then become 
\begin{subequations}
\begin{align}
    \lambda_i- s \sqrt{\lambda_i} > \mt_i^-, \\
    \lambda_i + s \sqrt{\lambda_i} < \mt_i^+.
\label{eq:lambda-sigma-ineq}
\end{align}
\end{subequations}
By solving these inequalities, we obtain the expression
\begin{align}
     \gamma t_i(\varepsilon,s) \geq \max{ \frac{1}{4} \Biggl( \hspace{-0.05cm} \sqrt{ \frac{ s^2 \abs{l_i}^2 }{ \Delta_\pm^2} + \frac{ 4X_\pm }{ \Delta_\pm} } \mp \frac{ s \abs{l_i} }{ \Delta_\pm } \Biggr)^{\hspace{-0.05cm}2}, 0},
\end{align}
for the convergence time given the error $\varepsilon$ and confidence level $s$. The parameters $\Delta_\pm$ and $X_\pm$ are defined as
\begin{align}
     	&\Delta_\pm = \frac{ \abs{l_{i\pm1}}^2-\abs{l_{i}}^2 + \abs{l_{i}}^2 \ln{ \frac{\abs{l_{i}}^2 }{ \abs{l_{i\pm1}}^2 } } }{ \ln{ \frac{\abs{l_{i}}^2 }{ \abs{l_{i\pm1}}^2 } } }, \\
	&X_\pm = \frac{\abs{\xt_\varepsilon} -\ln{\frac{p_i}{p_{i\pm1}}} }{  \ln{ \frac{\abs{l_{i}}^2 }{ \abs{l_{i\pm1}}^2 } } }.
\end{align}

\section{Discussion} \label{sec:generalisations}

\subsection{Dark states}\label{sec:darkstates}
A special case that deserves its own discussion is the presence of dark states. 
Here, we characterise a dark state by the inability to generate a quantum jump, and we inquire how these states manifest in our statistical analysis. Mathematically, this situation is represented by an eigenvalue of the jump operator $\L$ being 0. If this is the case,  our labelling convention tells us that $0=\abs{l_1}^2< \abs{l_2}^2 < ... < \abs{l_d}^2$, and the effects due to the dark state are obtained by taking the limit  $\abs{l_1}^2\to =0^+$ in the previous results. If the system is prepared with a non-zero probability of being in the dark state, one finds that
\begin{subequations}
\begin{align}
	&\lim_{t\to \infty} \Pt(1|m=0) = 1, \\
	&\Pt(1|m>0)  = 0, \quad \forall t, 
\end{align}
\end{subequations}
where the probability that $m=0$ in the long-time limit is given by $\displaystyle \lim_{t\to \infty} \Pt(m=0) = p_1$. 
While for all the other states, we instead find that
\begin{subequations}
\begin{align}
	&\lim_{t\to \infty} \Pt(i\neq 1|m=0) = 0, \\
	\label{eq:Pcond-non-dark}
	&\Pt(1|m>0)  = \frac{1}{1+\displaystyle \sum_{\substack{j\neq i \\ j\neq 1}} \exp{x_{ij}}}, \quad \forall t. 
\end{align}
\end{subequations}
In words, these results show that even a single jump acts as a switch that makes the dark state inaccessible for the rest of the conditional time evolution, whereas the absence of jumps over a long time instead projects the state into the dark state. 
The expression in \eqref{eq:Pcond-non-dark}, moreover, demonstrates that the conditional dynamics for all results $i\neq 1$ behave as if the dark state does not exist if at least one jump occurred, $m>0$.  Regarding the convergence rates and times, there are no significant changes except that the state $\ket{2}$ now becomes distinguishable from $\ket{1}$ at a rate that diverges logarithmically, while its overall convergence is still limited by the distinguishability from $\ket{3}$.

\subsection{Degeneracy}\label{sec:degeneracy}
In the following, we discuss the ramifications of the existence of degeneracies. If two or more eigenvalues coincide, the decomposition of $\L^\dagger \L$ in terms of its eigenstates no longer consists of only rank-1 projectors, but also contains higher-dimensional projectors onto the eigenspaces corresponding to the degenerate eigenvalues. In this case, the non-degeneracy condition in \secref{sec:non-degeneracy-condition} implies that a trajectory generated by the measurement (\ref{eq:weak-measurement}) cannot distinguish between eigenstates within the same eigenspace. In our statistical analysis, this indistinguishability is evident from the fact that degenerate eigenvalues will manifest as a single Poissonian mode in $\Pt(m)$, weighted by the sum of their initial probabilities. Moreover, letting I$_\lambda$ denote the index set for the degenerate eigenstates corresponding to the eigenvalue $\lambda$, one finds that the conditional probabilities for outcomes $i \in {\rm I}_\lambda$ are now upper-bounded according to
\begin{subequations}
\begin{align}
 	\Pt(i\in{\rm I}_\lambda |m) &= \frac{1}{1+\displaystyle \sum_{\substack{j\neq i :\\ \abs{l_j}^2 = \abs{l_i}^2}} \frac{p_j}{p_i} +\displaystyle \sum_{\substack{j\neq i :\\ \abs{l_j}^2 \neq \abs{l_i}^2}} \exp{x_{ij}}}, \\
 	&\leq \frac{p_i}{ \sum_{j\in {\rm I}_\lambda} p_j } < 1, \quad \text{if} \quad p_i \neq 1. 
\end{align}
\end{subequations}
The conditional dynamics, nevertheless, still distinguish between distinct eigenvalues, and the proof in the previous sections also holds for the degenerate case, if one associates the degenerate eigenspace with a single Poissonian mode. To encompass this more general situation, we label the distinct, but possibly degenerate, eigenvalues as $\abs{l_{\rm I}}^2$. In terms of these labels, we may write the probability distribution $\Pt(m)$ as
\begin{subequations}
\begin{equation}
	\Pt(m) = \sum_{\rm I } p_{\rm I} \, \Pois(\lambda_{\rm I};m), 
\end{equation}
where $p_{\rm I}$ and $\lambda_{\rm I} $ are defined as
\begin{align}
	&p_{\rm I} =\displaystyle  \sum_{i:\, \abs{l_i}^2 = \,\abs{l_{\rm I}}^2 } p_i,
 	&\lambda_{\rm I} = \gamma t \abs{l_I}^2.
\end{align}
\label{eq:Ptm-degenerate}
\end{subequations}
Similarly, we may write the corresponding conditional probabilities for obtaining the outcomes $\abs{l_{\rm I}}^2$ as
\begin{subequations}
\begin{align}
	\Pt( {\rm I} | m) &= \frac{ p_{\rm I}\, \abs{ l_{\rm I} }^{2m} \exp{- \gamma t \abs{l_{\rm I} }^{2} } }{ \displaystyle \sum_{\rm I } p_{\rm I}\, \abs{ l_{\rm I}}^{2m} \exp{-\gamma t \abs{l_{\rm I}}^{2}} }, \\ 
	& = \frac{ 1 }{ 1 + \displaystyle \sum_{\rm J \neq I} \exp{x_{\rm IJ}}},
\end{align}
\label{eq:Pcond-degenerate}
\end{subequations}
with $x_{\rm IJ}$ defined as in \eqref{eq:xij}, but with $\abs{l_{\rm I}}^2$ and $\abs{l_{\rm J}}^2$ now labelling eigenvalues that are distinct from each other, but each possibly degenerate. Equations (\ref{eq:Ptm-degenerate}) and (\ref{eq:Pcond-degenerate}) present the necessary objects to generalise our proof to the degenerate case, and the subsequent steps follow analogously. The implication for the asymptotic behaviour of the quantum trajectories is that each trajectory converges towards the projection of the initial state into one, randomly chosen, eigenspace corresponding to one of the eigenvalues $\abs{l_{\rm I}}^2$, where the dimension of this eigenspace is given by the multiplicity of $\abs{l_{\rm I}}^2$.

\subsection{The role of non-degenerate symmetry subspaces} \label{sec:strong-symmetry}
In Ref.~\cite{tindall_generality_2023}, the authors identified the existence of a strong symmetry as a condition allowing individual quantum-jump trajectories to ``freeze'' into single symmetry sectors of the symmetry. However, it was also noted that this phenomenon 
does not occur between similar symmetry subspaces, meaning that one symmetry subspace can be unitarily transformed into the other modulo a phase factor. In this section, we reassess these results with our operationally motivated approach, aiming to provide a physical interpretation.

In contrast to closed quantum systems, symmetries in open quantum systems may be defined in different ways, correspondingly describing different properties depending on the definition~\cite{albert_symmetries_2014}. Here, as in Ref.~\cite{tindall_generality_2023}, by symmetry we refer to a strong symmetry as defined in Ref.~\cite{buca_note_2012}. Such strong symmetry is characterised by a symmetry operator that is invariant under time evolution governed by the master equation. 
Formally, we call the operator $\hat{A}$ a strong symmetry if  
\begin{equation}
	\comm{\H}{\hat{A}}=\comm{\L_j}{\hat{A}} =\comm{\L_j^\dagger}{\hat{A}} = 0,
\end{equation}
for all Lindblad operators $\L_j$.

Here, we do not a priori assume there exists a strong symmetry. Nevertheless, our assumption that $\big[\H,\L\big]=0$ (condition (ii) in \secref{sec:implementing-a-projective-measurement}) implicitly identifies the jump operator $\L$ as a strong symmetry.
If $\L$ is diagonalisable in an orthonormal basis $\curly{\ket{i}}$, we may write $\L \ket{i} = l_i \ket{i}$, as before. Then, if the symmetry subspaces labelled by $i$ and $j$ are similar, we may also write $l_i = \exp{i\theta} l_j,$ with some phase $\theta \in \mathbb{R}$. Clearly, $\abs{l_i}^2 = \abs{l_j}^2$ in this case, which renders the subspaces $i$ or $j$ indistinguishable and thus prevents the asymptotic convergence to only one of them. Thereby, our operational perspective and non-degeneracy condition explain why similar symmetry subspaces may not permit asymptotic convergence to a single symmetry subspace.

The example above does, however, not yet provide the complete picture. In the case where $\L$ is diagonalisable and  $[\H,\L]=0$, we also find that $\L^\dagger \L$ is a strong symmetry. This is easy to check, as two diagonal operators always commute. In contrast to $\L$, the strong symmetry $\L^\dagger\L$ can, by construction, not have similar symmetry subspaces in the basis $\curly{\ket{i}}$, and we see that the non-degeneracy condition $\abs{l_i}^2 \neq \abs{l_j}^2$ is directly associated with non-degeneracy among its symmetry subspaces. 
In conclusion, for finite-size Hilbert spaces with a diagonalisable jump operator $\L$ that commutes with $\H$, we find that the measurement perspective identifies the measurement effect $\hat{M}_1^\dagger \hat{M}_1(dt) \propto \L^\dagger \L$, not the jump operator $\L$, as the relevant, \textit{observable} strong-symmetry operator that governs the asymptotic behaviour of individual quantum trajectories.

\subsection{Non-degeneracy condition for diffusive trajectories} \label{sec:diffusive-case}
So far, we have focused on a measurement process that manifests in stochastically occurring quantum jumps, with smooth evolution of the conditional state in between. In this case, there is an abrupt change in the observer's knowledge about the system when the detector clicks. 
One can, on the other hand, also conceive a measurement process where the information change is small but continuous in time. Such a monitoring process yields conditional state updates that describe \textit{diffusive} quantum trajectories. 
Operationally, diffusive quantum trajectories can be encountered when interfering a cavity mode's output field with a strong classical field (often referred to as a local oscillator) before detection. For example, they occur in homodyne or heterodyne detection, which are frequently employed measurement techniques in experimental quantum optics. In this situation, most of the detected signal stems from the classical control field, rendering the information gained about the system minimal. The outcome is a stochastic but smooth detection signal, rendering a diffusive quantum trajectory.

A diffusive quantum trajectory is mathematically described by a stochastic Wiener process. From the operational perspective taken above, such a Wiener process can be related to the quantum jump process described by the measurement (\ref{eq:weak-measurement}) by making the replacements
\begin{subequations}\label{eq:diffusive-replacement}
\begin{align} 
    & \L \to \L + \alpha, \\
    & \H \to \H - \tfrac{i\hbar}{2} (\alpha^*\L-\alpha\L^\dagger),
\end{align}
\end{subequations}
and taking the limit $\abs{\alpha} \to \infty$~\cite{wiseman_quantum_1993,wiseman_quantum_2010}. This replacement can be realised, e.g., through a beamsplitter interaction that interferes a cavity mode's output field with the local oscillator field whose field amplitude is proportional to $\alpha \in \mathbb{C}$. 

For our goal of finding the conditions for implementing a projective measurement in the limit of diffusive continuous monitoring, however, it is not necessary to make the formal transition to a Wiener process. Instead, we maintain our dynamical description in terms of quantum jumps, which is applicable for any finite $\alpha$, and take the limit of large $\alpha$ only after arriving at the convergence rates. This also reflects the actual experimental situation, where local oscillators are large but finite, and a description in terms of a Wiener process is merely due to the inability to resolve the quantum jumps. Moreover, this approach allows us to continuously interpolate towards the diffusive limit. We refer to the replacements in \eqref{eq:diffusive-replacement} with $\alpha$ finite as discrete homodyne detection~\cite{wiseman_quantum_1993}. We also note that the transformation of the Hamiltonian, which ensures the invariance of the Lindblad equation, is not required for our results, where different $\alpha$ can be associated with independent measurement protocols.

Making the same assumptions as above, that is, for a finite-dimensional quantum system, we assume $\L$ is diagonalisable and $\big[\H,\L\big]=0$. We see that mixing the system's output with a local field offsets the discrete eigenvalue spectrum of $\L$, as well as $\L^\dagger \L$, proportionally to $\alpha$. Thus, the corresponding non-degeneracy condition is readily obtained by replacing $l_i\to l_i+\alpha$ in \eqref{eq:non-degeneracy-condition}, leading to the non-degeneracy condition, 
\begin{equation}
	\abs{l_i+\alpha}^2 \neq \abs{l_j+\alpha}^2.
\end{equation}
If the phase of the local field is chosen such that $\alpha\in \mathbb{R}$, the condition above can be reduced to
\begin{subequations}\label{eq:diffusive-non-degeneracy-condition}
\begin{equation}\label{eq:X-non-degeneracy-condition}
	\Re{l_i} \neq \Re{l_j}.  
\end{equation}
This condition lifts any degeneracy among the previously unshifted eigenvalues $\abs{l_i}^2$ and ensures asymptotic convergence towards one of the eigenstates of $\L^\dagger \L$. 
On the other hand, if the phase is chosen such that  $\alpha$ is purely imaginary, we find that the sufficient condition is
\begin{equation}\label{eq:Y-non-degeneracy-condition}
	\Im{l_i} \neq \Im{l_j}.  
\end{equation}
These two approaches are often referred to as an $X$- and $Y$-quadrature measurement, respectively. In fact, the observable in this situation is $(\L+\alpha)^\dagger(\L+\alpha)$, rather than $\L^\dagger \L$, which contains information about the $X$- and $Y$-quadrature operators $(\L^\dagger+\L)$ and $i(\L^\dagger - \L)$ when $\alpha$ is, respectively, real and purely imaginary. 
The third situation would be a heterodyne detection scheme, which simultaneously probes the $X$- and $Y$-quadratures. In this case, satisfying either of the conditions (\ref{eq:X-non-degeneracy-condition}), or (\ref{eq:Y-non-degeneracy-condition}) is sufficient. 
\end{subequations}

Equation\,(\ref{eq:X-non-degeneracy-condition}) reproduces the non-degeneracy condition found necessary for the asymptotic convergence of individual diffusive trajectories in Ref.~\cite{benoist_large_2014}. In contrast to~\cite{benoist_large_2014}, however, we derive the condition for the diffusive case directly from the quantum-jump condition. As such, our results demonstrate that the diffusive and quantum-jump non-degeneracy conditions emerge from the same operational principle, where the observer's role in ascribing physical meaning to the quantum trajectory is central.

\subsubsection{Convergence rate for diffusive quantum trajectories}
It is also interesting to see how the offset by $\alpha$ affects the convergence rate. As for quantum-jump trajectories, we identify the convergence rate as 
\begin{subequations}
\begin{align}
	R_i^\alpha  = {\rm min} \Bigg\{ &\abs{l_{i\pm1}+\alpha}^2 \hspace{-0.09cm} -\abs{l_{i}+\alpha}^2 \nonumber \\
	& + \abs{l_{i}+\alpha}^2 \ln{\frac{\abs{l_{i}+\alpha}^2}{\abs{l_{i\pm1}+\alpha}^2}} \Bigg\},
\end{align}
\end{subequations}
given by making the replacements $l_i \to l_i+\alpha$ in \eqref{eq:convergence-rate}.
The effect of adding $\alpha$ is a sharp reduction of the convergence rate with increasing magnitude $\abs{\alpha}$, before it saturates at a constant value for $\abs{\alpha}\gg1$, corresponding to the convergence rate for diffusive quantum trajectories. An analytical expression for the saturated rate can be found by taking the limit $\abs{\alpha} \to \infty$. The details of this calculation are given in \appref{app:Kijk-diffusive} and the result for a general $\alpha=\abs{\alpha}\exp{\phi_\alpha}\in\mathbb{C}$ is 
\begin{widetext}
\begin{align}
	\label{eq:Ri-diffusive}
    \lim_{\abs{\alpha}\to \infty} R_i^{\alpha} = & \hspace{0.06cm} 
    {\rm min} \Bigg\{ 2 \hard{ \round{ \rel{l_{i\pm1}} \cos{\phi_\alpha}+\iml{l_{i\pm1}} \sin{\phi_\alpha} }^2  - \round{ \rel{l_i} \cos{\phi_\alpha}  +  \iml{l_i}\sin{\phi_\alpha} }^2  } \nonumber \\  
     & - 4\hard{\rel{l_i}\cos(\phi_\alpha) + \iml{l_i}\sin(\phi_\alpha)} 
     \hard{ \round{\rel{l_{i\pm1}}-\rel{l_i} }\cos(\phi_\alpha) +  \round{ \iml{l_{i\pm1}} - \iml{l_i}} \sin(\phi_\alpha)} \Bigg\},
\end{align}
\end{widetext}
which reduces to the following expressions for homodyne $X$ and $Y$ measurements, respectively, 
\begin{equation}
	 \lim_{\abs{\alpha}\to \infty} \abs{K_{ij,i}^{\alpha} }\hspace{-0.05cm} = \hspace{-0.05cm}
	 \begin{cases}
	 2(\rel{l_i}-\rel{l_{i\pm1}})^2, \;\;\; \alpha \in\mathbb{R}, \\
	2(\iml{l_i}-\iml{l_{i\pm1}})^2, \;\;\; \alpha = i \alpha,
	 \end{cases}
\end{equation}
It is now interesting to compare the result for the homodyne $X$ measurement with the diffusive convergence rate found in Ref.\,\cite{benoist_large_2014}. As in the quantum jump case, our approach, in contrast to \,\cite{benoist_large_2014}, allows us to identify the competition between neighbouring modes as the slowest rates.  Otherwise, our rate has the same relationship between the eigenvalues, but is, curiously, found to be 2 times larger than the one in Ref.\,\cite{benoist_large_2014}.

\subsection{Asymptotic convergence in the fixed-\texorpdfstring{$m$}{m} protocol} \label{sec:fixed-m-perspective}

We now turn to the complementary perspective in which the jump count $m$ is fixed and the arrival time $t$ becomes the random variable. This offers a distinct route to analysing the same measurement dynamics in terms of the first-passage-time statistics. In this perspective, $m$ takes the role of the asymptotic variable and, as we will outline below, the proof to show asymptotic convergence to individual eigenstates of the observable $\L^\dagger \L$ follows analogously to the proof in the fixed-$t$ perspective, provided in \secref{sec:asymptotic-convergence}.

The first-passage-time distribution is readily derived from the jump-count distribution using the identity $\Pm(t)= \partial t \sum_{k\geq m} \Pt(k)$~\cite{menczel_full_2026}. Employing \eqref{eq:Ptm} for $\Pt(m)$, we find that the first-passage-time distribution can be written as a weighted sum of Erlang distributions, analogous to the Poissonian modes in \eqref{eq:Ptm}:
\begin{subequations}
	\begin{align}\label{eq:Pmt}
		\Pm(t) 	&= \sum_{i=1}^d p_i \frac{ t^{m-1} \round{\gamma \abs{l_i}^2 }^m }{(m-1)!} \exp{-\gamma t \abs{l_i}^2} \\
				&= \sum_{i=1}^d p_i \Erl(m,\Gamma_i;t).
	\end{align}
\end{subequations}
The Erlang distribution is characterised both by its shape parameter, which here is the (fixed) jump count $m$, and the rate $\Gamma_i=\gamma \abs{l_i}^2$, while the corresponding mean of the $i$th mode is given by $\mu_i=m/\Gamma_i$.

\begin{figure}
    \centering
    \includegraphics[width=0.85\linewidth]{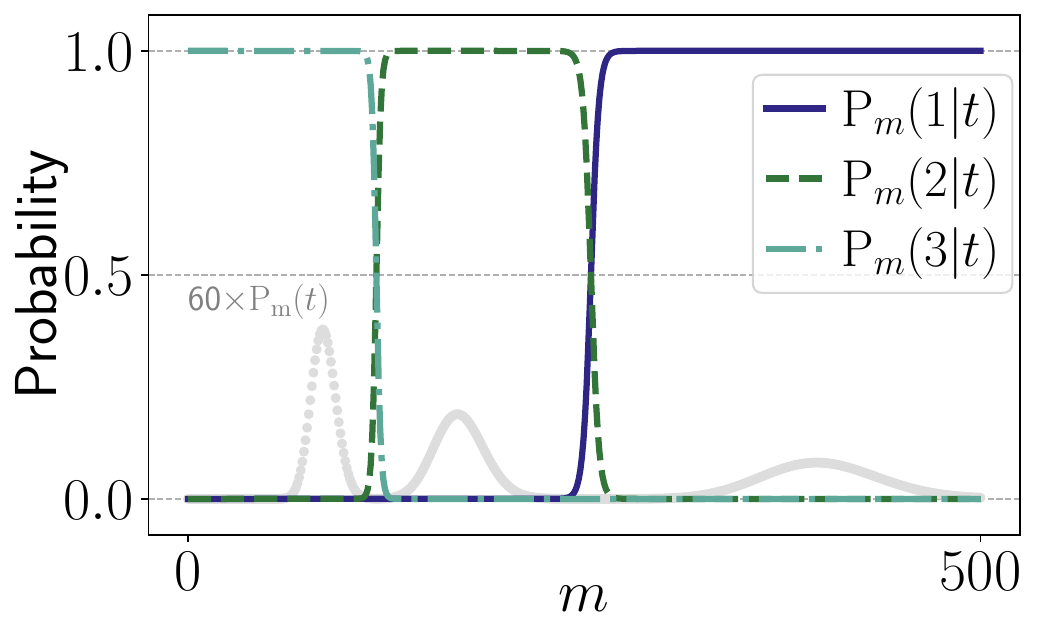}
    \caption{Illustration of the first-passage-time distribution $\Pm(t)$ and its relationship to the conditional probabilities $\Pm(i|t)$.  The eigenvalues and initial probabilities are the same as in \figpanel{fig:Proof-illustration}{b} and the jump count $m=120$ is chosen to match $t_{\rm long}$ in that figure.}  
    \label{fig:fixed-m}
\end{figure}

Similar to the fixed-$t$ protocol, the first-passage-time distribution represents the marginal distribution of a joint distribution for the first-passage time $t$ and the outcome $\abs{l_i}^2$, obtained as the result of a hypothetical, ideal projective measurement on the conditional state performed right after the arrival at a preset jump count $m$. Accordingly, we derive the conditional probability distribution for the outcome $\abs{l_i}^2$ given the first-passage time $t$ as
\begin{subequations}
\begin{align} \label{eq:Pcond-m}
	\Pm(i | t) =  \frac{ p_i \abs{l_i}^{2m} \exp{-\gamma t \abs{l_i}^2 } }{ \displaystyle \sum^d_{i=1} p_i \abs{l_i}^{2m} \exp{-\gamma t \abs{l_i}^2 } },
\end{align}
where we again employ the simpler label $i$ for the outcome $\abs{l_i}^2$. Notably, the expression above is identical to the one in \eqref{eq:Ptcond}. The difference lies in the roles played by its input variables. Here, the jump count $m$ is the reference variable, whose asymptotic limit we wish to investigate, and the time $t$ represents the first passage time instead of an externally controlled measurement time. 
Since the expressions are otherwise identical, we may rewrite \eqref{eq:Pcond-m} in terms of the variables $x_{ij}$ defined in \eqref{eq:xij}, which yields
\begin{align}
	\Pm(i | t) =  \frac{ 1 }{ 1 + \displaystyle \sum_{j \neq i} \exp{x_{ij}} }.
\end{align}
\end{subequations}
From this point, the proof follows analogously to the steps taken from \eqref{eq:Ptcond_xij}. Specifically, we recast the proof in terms of the parameter $\tfrac{\gamma t}{m}$, which is identified as the reciprocal of $\tfrac{m}{\gamma t}$. This allows us to interpret $\Pm(i | t)$ as the reciprocal of $\Pt(i|m)$, and we may redraft the illustrative picture provided in \figpanel{fig:Proof-illustration}{d} in terms of $\Pm(i | t)$ as a function of $\tfrac{\gamma t}{m}$, fully analogously. The technical details are provided in \appref{sec:app-fixed-m-perspective}. In \figref{fig:fixed-m}, we also illustrate the mode structure of $\Pm( t)$ for a large $m$ and its association to the distinct intervals where the conditional probabilities $\Pm(i|t)=1$, analogous to \figpanel{fig:Proof-illustration}{b}.

To conclude, our results show that the first-passage-time statistics generated by the continuous measurement process (\ref{eq:weak-measurement}), similar to the counting statistics, asymptotically demonstrate the implementation of an ideal projective measurement of the observable $\L^\dagger \L$ when $[\H,\L] =0$ and the non-degeneracy condition (\ref{eq:non-degeneracy-condition}) is fulfilled $\forall i$. 
Our results additionally highlight the first-passage-time statistics as an alternative and viable route for analysing the conditional dynamics of individual quantum trajectories in a continuous measurement setting.

\subsubsection{Convergence rate and finite-\texorpdfstring{$m$}{m}-error  relationship} 
As in the fixed-$t$ case, we may also inquire what the minimum jump count is to ascribe the state $\ket{i}$ to the conditional state with a small but finite error $\varepsilon$. By requiring that $\Pm(i|t)\geq 1-\varepsilon$, we derive the threshold times
\begin{equation} \label{eq:threshold-times}
	\gamma \tilde{t}_i^\pm = \frac{m \ln{\frac{\abs{l_i}^2}{\abs{l_{i\pm1}}^2}} + \xt_\varepsilon + \ln{\frac{p_i}{p_{i\pm1}}} }{  \abs{l_i}^2 - \abs{l_{i\pm1}}^2}
\end{equation}
bounding the interval $\tilde{t}_i^+ \leq t \leq \tilde{t}_i^-$ within which the bound on the conditional probability stated above is satisfied. The parameter $\xt_\varepsilon$, depending on the error $\varepsilon$, is given by \eqref{eq:xep} as before.

The times in \eqref{eq:threshold-times} also define the inequalities $\tilde{t}_i^+ < \mu_i < \tilde{t}_i^-$ which the mean $\mu_i$ of the $i$th mode in $\Pm(t)$ must simultaneously satisfy to achieve on average the precision $\varepsilon$. Consequently, we obtain the minimum jump count $m_i(\varepsilon)$ by solving the two inequalities and maximising the two results:
\begin{equation} \label{eq:m-min-finite-ep}
	m_i(\varepsilon) >  \max{\frac{\abs{\xt_\varepsilon}-\ln{\frac{p_i}{p_{i\pm1}} } }{\ln{\frac{\abs{l_i}^2}{\abs{l_{i\pm1}}^2} } -\frac{\round{\abs{l_i}^2-\abs{l_{i\pm1}}^2} }{\abs{l_i}^2 } } }.
\end{equation}
From this expression, we may additionally identify the corresponding convergence rate
\begin{equation}\label{eq:rate-fixed-m}
	R^m_i = 	\min{ -\frac{ 1 }{ \abs{l_i}^2 } \round{ \abs{l_{i\pm1}}^2 - \abs{l_i}^2 + \ln{\frac{\abs{l_i}^2}{\abs{l_{i\pm1}}^2} } } },
\end{equation}
where we assume that $|l_i|^2 \neq 0$ (no dark state). Comparing this rate with the convergence rate in \eqref{eq:convergence-rate}, we see that it may be faster or slower, depending on the explicit values of the eigenvalues of $\L^\dagger \L$. Consequently, a judicious choice of the protocol, fixed-$t$ or fixed-$m$, may achieve a higher statistical accuracy more efficiently.

\section{Examples} \label{sec:examples}

\subsection{Qubit \texorpdfstring{$\hat{\sigma}_z$ measurement}{sz measurement}}\label{sec:qubit-sz-measurement}
An instructive example is a two-level system or qubit subject to a non-destructive measurement of the Pauli operator $\hat{\sigma}_z$, where the internal dynamics are described by a Hamiltonian that satisfies $[\H,\hat{\sigma}_z]=0$. This example serves to demonstrate how the discussed quantum-jump and diffusive measurement schemes can result in entirely different convergence behaviour at the level of individual trajectories, while both give rise to the same ensemble average. 

In the quantum-jump measurement scheme, all trajectories return to the initial state on all even jump counts and thus fail to converge towards an eigenstate of $\hat{\sigma}_z$. Thus, the quantum-jump trajectory is unable to realise a projective measurement and, regarding its thermodynamic properties, the trajectory is ergodic as the time-average conditional state equals the ensemble-average state. In the diffusive measurement scheme, in contrast, each trajectory converges to one of the eigenstates of $\hat{\sigma}_z$ for sufficiently long measurement times. That is, each trajectory asymptotically realises a projective measurement of $\hat{\sigma}_z$ and is therefore not ergodic.

The two different behaviours are easily explained by our statistical analysis in the previous sections. In this case, the quantum jump operator is $\L=\hat{\sigma}_z$, which has the eigenvalues $l_0=-1$ and $l_1=1$.  However, the corresponding measurement effect is proportional to $\L^\dagger \L = \hat{\sigma}_z^2 = \hat{\mathbb{I}} $, which is fully degenerate. The quantum-jump non-degeneracy condition (\ref{eq:non-degeneracy-condition}) is thus not fulfilled. Moreover, by inserting $\abs{l_0}^2 = \abs{l_1}^2=1$ into Eqs.\,(\ref{eq:Ptm}) and (\ref{eq:Ptcond}), one finds that the distribution of quantum jumps $\Pt(m)$ is a single Poisson distribution with mean $\gamma t $, and that the conditional probabilities are fully determined by the initial conditions: $\Pt(i|m)=p_i= \mel{i}{\rhoh(0)}{i}$, $i=0,1$. Hence, it is clear that the quantum-jump measurement scheme characterised by the jump operator $\L=\hat{\sigma}_z$ cannot realise a projective measurement of $\hat{\sigma}_z$. 

The diffusive measurement scheme corresponding to the same jump operator, on the other hand, can. This is explained by homodyning the system's output signal. That is, it is interfered with a strong field from a local oscillator. As discussed in \secref{sec:diffusive-case}, this lifts any degeneracies among the eigenvalues of the measurement effect and leads to the diffusive non-degeneracy condition (\ref{eq:diffusive-non-degeneracy-condition}) that, in this example, is satisfied by $l_0$ and $l_1$. For comparison with the quantum-jump scheme, we consider what we referred to as discrete homodyne detection in \secref{sec:diffusive-case}, which does not require us to take the continuum limit. The interference with the local field with amplitude $\propto \alpha$ leads to a constant shift $l_i \to l_i + \alpha$. Consequently, the previously single Poissonian in $\Pt(m)$ is split into two modes with means $\lambda_0 = \gamma t \abs{\alpha -1}^2$ and $\lambda_1 = \gamma t \abs{\alpha +1}^2$ and the conditional probabilities for the corresponding outcomes $\abs{\alpha \pm 1}^2$ are described by \eqref{eq:Ptcond}. Furthermore, since the eigenvalues $\abs{\alpha \pm 1}^2$ satisfy the non-degeneracy condition, all subsequent steps of our proof are also satisfied. Hence, it follows directly that the diffusive measurement scheme implements a projective measurement of $\hat{\sigma}_z$ in the long-time limit.  

\begin{figure*}
    \centering
    \includegraphics[width=\linewidth]{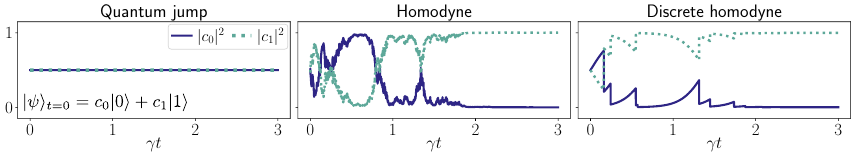}
    \caption{Sample quantum trajectories demonstrating the conditional state dynamics during a continuous measurement of $\hat{\sigma}_z$ on a qubit implemented by: quantum jump ($\L=\hat{\sigma}_z$, point process), homodyne ($\L=\hat{\sigma}_z$, Wiener process), and discrete homodyne ($\L=\hat{\sigma}_z + \alpha$, $\alpha$ finite, point process) detection. In the latter, we used $\alpha=1.8 \gamma$ and confirmed that the time step was small enough to ensure at most one detection per $[t,t+dt]$.} 
    \label{fig:examples}
\end{figure*}

In \figref{fig:examples}, we present sample trajectories of the three scenarios: quantum jump, homodyne, and discrete homodyne detection.

\subsection{Two coupled spin-\texorpdfstring{$\tfrac{1}{2}$}{1/2} particles}
As a second example, which exhibits both degeneracy and dark states, we consider two coupled spin-$\tfrac{1}{2}$ particles undergoing a continuous measurement of the collective spin polarisation. The Hamiltonian for the coupled spins is
\begin{equation}
 	\H = \omega \round{\Sx_1 \Sx_2 + \Sy_1\Sy_2 + \Sz_1 \Sz_2},
\end{equation}
where $\S^\alpha_1=\hat{\sigma}_\alpha \otimes \hat{\mathbb{I}}$ and $\S^\alpha_2 =\hat{\mathbb{I}} \otimes \hat{\sigma}_\alpha $, are local spin operators addressing each of the two coupled spins in their joint Hilbert space, and the $\hat{\sigma}_\alpha$, $\alpha=\curly{x,y,z}$, are Pauli operators.  
The jump operator corresponding to the continuous measurement is $\L=\S_z=\Sz_1+\Sz_2$, which has three distinct eigenvalues $l_0=0$, $l_1=1$ and $l_{-1}=-1$, where $l_0$ is twofold degenerate, and its subspace is spanned by dark states. Similar to the qubit case, the measurement effect, which is $\propto \S_z^2$, loses the information about the sign difference between $l_1$ and $l_{-1}$ and has instead two distinct eigenvalues $\abs{l_0}^2=0$ and $\abs{l_1}^2=1$ that are both twofold degenerate. 

The twofold degeneracy of $\abs{l_0}^2$ and $\abs{l_1}^2$ fails the non-degeneracy condition and prevents the quantum-jump trajectory from implementing a projection onto a single eigenstate. However, in this case, the trajectory \textit{can} distinguish between $\abs{l_0}^2$ and $\abs{l_1}^2$ and will, in the long-time limit, converge to \textit{one} of the respective eigenspaces, chosen at random. This is proven by taking $\abs{l_{\rm I}}^2 = \{ \abs{l_0}^2, \abs{l_1}^2 \}$ in our proof, employing the generalisation to degenerate subspaces outlined in \secref{sec:degeneracy}. Accordingly, we find that the quantum-jump trajectory realises a projection onto one of the two two-dimensional subspaces belonging to $\abs{l_0}^2$ and $\abs{l_1}^2$, but not onto a single eigenstate. Moreover, since the eigenspace of $\abs{l_0}^2$ is dark, the jump statistics in this measurement scheme are particularly straightforward to interpret: a single jump already projects the trajectory onto the bright subspace, and the probability for no jump to occur is determined by the initial probability for the state to be in the dark subspace.

As in the qubit example, we can also gain intuition for the behaviour of the analogue diffusive quantum trajectories by taking the perspective of discrete homodyning. The field of the local oscillator lifts the degeneracy between $\abs{l_1}^2$ and $\abs{l_{-1}}^2$ which now becomes $\abs{\alpha\pm1}^2$, respectively, and the dark subspace becomes bright in the sense that its is now statistically represented by a Poissonian mode in $\Pt(m)$ with mean $\gamma t \abs{\alpha}^2$. The twofold degeneracy of the dark subspace is, however, not lifted.  Accordingly, the diffusive trajectories distinguish the eigenspaces belonging to $l_0$, $l_1$, and $ l _ {- 1} $, and realise ideal projective measurements of the eigenstates $\ket{1}$ and $\ket{-1}$, respectively, while a trajectory that converges to the $l_0$ eigenspace implements a projection into the corresponding two-dimensional subspace.

\section{Conclusion}
We have established the conditions under which continuous monitoring, in the form of quantum-jump trajectories, gives rise to an ideal projective measurement. For finite-dimensional systems with a single, diagonalisable jump operator commuting with the Hamiltonian, we showed, based on the full counting statistics and the first passage time distribution, respectively, that individual trajectories asymptotically localise onto an eigenstate of the measurement operator whenever the corresponding eigenvalues of $\L^\dagger \L$ are non-degenerate. This non-degeneracy condition is not merely a technical requirement but reflects a deeper structural fact: it is equivalent to the absence of similar symmetry subspaces under the strong symmetry generated by $\L^\dagger \L$, linking the emergence of projective measurements directly to the symmetry structure of the underlying dynamics.

Beyond establishing convergence, we derived explicit expressions for the convergence rate and the associated time-error relationship, showing that the rate is set by the distinguishability between neighbouring eigenvalues rather than by the full spectrum, and that finite measurement times carry a quantifiable, tunable confidence level. We further extended the framework to account for degenerate eigenvalues and dark states, showing that trajectories still converge, but onto eigenspaces rather than individual eigenstates when degeneracies persist. Finally, by relating the quantum-jump case to diffusive (homodyne and heterodyne) measurement schemes through displacement of the jump operator, we recovered and generalised previously known non-degeneracy conditions for diffusive trajectories, unifying the two pictures under a single operational principle.

Taken together, these results provide a transparent and practically useful criterion for when and how quickly continuous monitoring can be trusted to realise an ideal projective measurement, with direct relevance to the design and optimisation of measurement protocols in quantum information experiments.

A natural next step is to extend this framework to multiple jump operators and to infinite-dimensional systems, where the interplay between several strong symmetries, or a continuous spectrum, may qualitatively change the convergence behaviour. Equally important for practical settings is to understand the robustness of these results: how small perturbations, unmonitored loss channels, and imperfect detection efficiency affect the asymptotic convergence and its rate. 

Finally, relaxing the assumption that the jump operator and the Hamiltonian commute would allow the framework to address a substantially broader class of physically realistic measurement scenarios, where coherent dynamics compete with the measurement back-action. Addressing these questions would extend the applicability of our results beyond the idealised setting considered here and bring the framework closer to realistic experimental conditions.

\section{Acknowledgements}
T. K. performed this work as an International Research Fellow of the Japan Society for the Promotion of Science (JSPS). 
F. N. is supported in part by the Japan Science and Technology Agency (JST)
[via the CREST Quantum Frontiers program Grant No. JPMJCR24I2], the Quantum Leap Flagship Program (Q-LEAP), the Moonshot R\&D Grant Number JPMJMS256E, and the ASPIRE program (Grant Number JPMJAP2513)].

\appendix

\section{Determining the dominant exponentials}\label{sec:app-xmax}
\begin{figure}
    \centering
    \includegraphics[width=\linewidth]{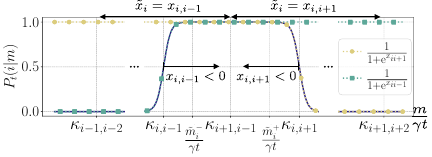}
    \caption{Illustration of the relationships found in Eqs.\,(\ref{eq:app-xmax1}) and (\ref{eq:app-xmax2}). The bounding jump counts $\mt_i^{\pm}$ [\eqref{eq:mipm}], between which $\Pt(i|m)\geq 1-\varepsilon$, are also indicated. In this figure, $\varepsilon=10^{-3}$.}
    \label{fig:app-ptcond-intervals}
\end{figure}

In this appendix, we determine $\xt_i=\max{x_{ij}}_{j\neq i}$ in the interval where $\Pt(i|m)\approx 1$. For an easier reading, we repeat the definitions of $x_{ij}$ and $\kappa_{ij}$, introduced in Eqs.\,(\ref{eq:xij}) and (\ref{eq:kij}), respectively:
\begin{equation}
	 x_{ij} = \gamma t \round{ \abs{l_i}^2 -\abs{l_j}^2 } - m \ln{\frac{\abs{l_i}^2}{\abs{l_j}^2}} - \ln{\frac{p_i}{p_j}},
\end{equation}
\begin{equation}
\label{eq:app-kij}
	\kappa_{ij} = \frac{ \abs{l_i}^2 -\abs{l_j}^2 }{ \ln{ \frac{\abs{l_i}^2}{\abs{l_j}^2} } }.
\end{equation}
We begin by noting a few useful properties. First, we note that 
\begin{subequations}
\begin{equation}
\label{eq:app-prop1}
	\kappa_{ij}=\kappa_{ji},
\end{equation}
and that
\begin{equation}
\label{eq:app-prop2}
	\kappa_{ij}>\kappa_{ij'}, \; \; \text{if} \;\; \abs{l_j}^2>\abs{l_{j'}}^2.
\end{equation}
The second property is easily seen  by defining
$x\coloneq \frac{\abs{l_i}^2}{\abs{l_j}^2} $ and $x' \coloneq \frac{\abs{l_i}^2}{\abs{l_{j'}}^2}$, which allows us to write
\begin{align}
	\frac{\kappa_{ij}}{\kappa_{ij'}} &= \frac{1-\frac{1}{x}}{\ln{x}} \frac{\ln{x'}}{1-\frac{1}{x'}}>1, \; \quad \text{if} \; \; x < x'. \nonumber
\end{align} 
Thus, we find that 
\begin{align}
	\frac{\kappa_{ij}}{\kappa_{ij'}} >1, \; \quad \text{if} \; \; \abs{l_j}^2 > \abs{l_{j'}}^2. \nonumber
\end{align} 
The third property is
\begin{align}
\label{eq:app-prop3}
	x_{ij}>x_{ij'} \quad &\text{when} \quad \frac{m}{\gamma t} \lessgtr \kappa_{jj'}, \\
	\text{if} \quad \abs{l_j}^2 \lessgtr \abs{l_{j'}}^2, \quad &\text{and} \quad \gamma t \gg \frac{\ln{\frac{p_{j'}}{p_j}}}{\ln{\frac{\abs{l_{j'}}^2}{\abs{l_j}^2}}}. \nonumber
\end{align}
To derive this, we first note that
\begin{align}
	x_{ij}-x_{ij'} = x_{j'j}. \nonumber
\end{align}
Thus, we find that 
\begin{align}
	x_{ij}>x_{ij'} \;\; \Leftrightarrow \;\; x_{j'j}>0, \nonumber
\end{align} 
and
\begin{align}
	 &x_{j'j}>0. \nonumber \\
	 &\gamma t \round{\abs{l_{j'}}^2-\abs{l_j}^2} - m\ln{\frac{\abs{l_{j'}}^2}{\abs{l_{j}}^2}} - \ln{\frac{p_{j'}}{p_j}} > 0\nonumber \\
	 &\frac{m}{\gamma t} \lessgtr \frac{\round{\abs{l_{j'}}^2-\abs{l_j}^2} }{\ln{\frac{\abs{l_{j'}}^2}{\abs{l_j}^2}}} - \frac{1}{\gamma t} \frac{\ln{\frac{p_{j'}}{p_j}}}{\ln{\frac{\abs{l_{j'}}^2}{\abs{l_j}^2}}}  , \quad \abs{l_j}^2 \lessgtr \abs{l_{j'}}^2. \nonumber
\end{align} 
Equation (\ref{eq:app-prop3}) follows directly for $\gamma t \gg  \ln{\frac{p_{j'}}{p_j}} / \ln{\frac{\abs{l_{j'}}^2}{\abs{l_j}^2}} $.
From Eqs.\,(\ref{eq:app-prop2}) and (\ref{eq:app-prop3}), it also follows that 
\begin{align}
	\label{eq:app-prop4}
	& \max{x_{ij_1}, x_{ij_2}, x_{ij_3} ...}=x_{ij_1}, \\
	& \text{when} \; \frac{m}{\gamma t} < \kappa_{j_1j_2} < \kappa_{j_1j_3} < ... , \nonumber \\
	& \text{if} \;\; \abs{l_{j_1}}^2 < \abs{l_{j_2}}^2 < \abs{l_{j_3}}^2 < ... \, ,\nonumber 
\end{align}
and
\begin{align}
	\label{eq:app-prop5}
	& \max{x_{ij'_1}, x_{ij'_2}, x_{ij'_3} ...}=x_{ij'_1}, \\
	& \text{when} \; \frac{m}{\gamma t} > \kappa_{j'_1j'_2} > \kappa_{j'_1j'_3} > ... \, , \nonumber \\
	& \text{if} \;\; \abs{l_{j'_1}}^2>\abs{l_{j'_2}}^2>\abs{l_{j'_3}}^2 >...\, . \nonumber
\end{align}
\end{subequations}
Equipped with properties (\ref{eq:app-prop1})-(\ref{eq:app-prop5}), we now move on to determine $\xt_i$. The results we obtain in the following are illustrated schematically in \figref{fig:app-ptcond-intervals}. 

First, let us remind the reader that we sort the eigenvalues of $\L^\dagger \L$ such that $\abs{l_1}^2 < \abs{l_2}^2 < ... < \abs{l_d}^2 $, as in the main text.
Secondly, based on the argument of statistical indistinguishability, it is a reasonable guess that $\xt_i=x_{i,i-1}$ or $x_{i,i+1}$ in the interval where $\Pt(i|m) \approx 1$. Therefore, we begin by noting that  
 \begin{subequations}
 \begin{align}
 	&x_{i,i-1} > x_{i,i+1} \; \; \text{for} \; \; \frac{m}{\gamma t} < \kappa_{i-1,i+1}, \\
	&x_{i,i+1} > x_{i,i-1} \; \; \text{for} \; \; \frac{m}{\gamma t} > \kappa_{i-1,i+1}.
 \end{align} 
\end{subequations}
Next, we focus on $x_{i,i-1}$ and label the eigenvalues such that $\abs{l_{i,i-1}}^2 < \abs{l_{i,i+1}}^2 < \abs{l_{i,i+2}}^2 < ... $ and $\abs{l_{i,i-1}}^2 > \abs{l_{i,i-2}}^2 > \abs{l_{i,i-3}}^2 > ... $ . Then,  for large $\gamma t$,  Eqs.\,(\ref{eq:app-prop4}) and (\ref{eq:app-prop5}) tells us that 
\begin{align}
	&\max{x_{i,i-1},x_{i,i+1}, x_{i,i+2}, ...} = x_{i,i-1}, \\
	&\text{for}\; \; \frac{m}{\gamma t} < \kappa_{i-1,i+1} < \kappa_{i-1,i+2} < ...\, , \nonumber
\end{align}
and
\begin{align}
	&\max{x_{i,i-1},x_{i,i-2}, x_{i,i-3}, ...} = x_{i,i-1}, \\
	& \text{for}\; \; \frac{m}{\gamma t} > \kappa_{i-1,i-2} > \kappa_{i-1,i-3} > ...\, . \nonumber
\end{align}
Moreover, according to \eqref{eq:app-prop2}, we find that $\kappa_{i-1,i+1}>\kappa_{i-1,i-2}$. Taken together, these results imply
\begin{equation}
	\xt_i=x_{i,i-1} \; \; \text{when} \;\; \kappa_{i-1,i-2} < \tfrac{m}{\gamma t} < \kappa_{i-1,i+1}.
\end{equation}
However, we also require that $\xt_{i}<0$ so that $\exp{\xt_i} \approx 0$ for $\abs{\xt_i} \gg 1$. Solving $x_{i,i-1}<0$ gives the condition $\tfrac{m}{\gamma t}> \kappa_{i,i-1}$ and, employing \eqref{eq:app-prop2}, we find that $\kappa_{i,i-1}>\kappa_{i-1,i-2}$. To conclude, these results demonstrate that 
\begin{align}
	\label{eq:app-xmax1}
	&\xt_i=x_{i,i-1} \;\; \text{and} \;\; x_{i,i-1} < 0, \\
	&\text{when} \;\; \kappa_{i,i-1} < \tfrac{m}{\gamma t} < \kappa_{i-1,i+1}. \nonumber
\end{align}
Analogous calculations for $x_{i,i+1}$ provides the complementary result 
\begin{align}
	\label{eq:app-xmax2}
	&\xt_i=x_{i,i+1} \;\; \text{and} \;\; x_{i,i+1} < 0, \\
	&\text{when} \;\; \kappa_{i-1,i+1} < \tfrac{m}{\gamma t} < \kappa_{i,i+1}. \nonumber
\end{align}

\section{Cumulative tail probabilities} \label{app:vanishing-tails}
The goal of this appendix is to show that the bounds on the tail probabilities of the Poisson distributions, presented in \eqref{eq:tail-bounds}, vanish in the limit $t \to \infty$. For clarity, we repeat these bounds here, which are 
\begin{subequations}
\begin{align}
    \Pois\round{\lambda_k; m \geq  \mt_{i}^+ } \leq \frac{({\rm e} {\lambda_k})^{ \mt_{i}^+ }\exp{-\lambda_k}}{(\mt_{i}^+)^{\mt_{i}^+}} ,
\end{align}
when  $\lambda_k <\mt_{i}^+$, and
\begin{align}
   \Pois\round{\lambda_k; \hspace{-0.03cm} m  \hspace{-0.03cm}\leq  \hspace{-0.03cm}\mt_{i+1}^-  } \hspace{-0.07cm}\leq\hspace{-0.07cm} \frac{({\rm e} {\lambda_k})^{{\mt_{i+1}^-} }\exp{-\lambda_k}}{ ({\mt_{i+1}^-})^{\mt_{i+1}^-}},
\end{align}
\label{eq:app-poisson-bounds}
when $\lambda_k  >  \mt_{i+1}^-$.
\end{subequations}
The index $i\in(1,d-1)$ labels the $d-1$ crossings where $\Pt(i|m)$ gives way for $\Pt(i+1|m)$. We additionally want to show that $\lambda_k < \mt_i^+$, $\forall \lambda_k \leq \lambda_i$ and $\lambda_k > \mt_{i+1}^-$, $\forall \lambda_k \geq \lambda_{i+1}$ for long measurement times $ t$ so that the vanishing of the bounds above implies that $\lim_{t\to \infty} \sum_{i=1}^{d-1} \Pt(\mt_i^+ \leq m \leq \mt_{i+1}^- ) = 0$.

\subsection{Characterising the \texorpdfstring{$\lambda_k$}{Lk}} \label{app:characterising-lambda}
Beginning with the second task, we recall that $\lambda_k = \gamma t \abs{l_k}^2$ and use \eqref{eq:mipm} for $\mt_i^+$ and $\mt_{i+1}^-$ to evaluate the inequalities. 
By defining the parameters $c=\tfrac{\abs{l_{i}}^2}{\abs{l_{k}}^2}$, and $u = \tfrac{\abs{l_{i+1}}^2}{\abs{l_{i}}^2}$, the first inequality can be rewritten as %
\begin{subequations}
\begin{align}
	\label{eq:app-meanvalue-1a}
	&\lambda_k< \mt_i^+,  \\
	\label{eq:app-meanvalue-1b}
	&\ln{\frac{1}{u}} < c\round{1-u}  + \frac{\abs{\xt_{\varepsilon}} - \ln{\frac{p_i}{p_{i+1}}}}{\gamma t \abs{l_k}^2}. 
\end{align}
\end{subequations}
Since $\abs{\xt_\varepsilon}$ is finite, albeit arbitrarily large corresponding to an arbitrarily small error $\varepsilon$, the last term on the second line may be neglected for sufficiently long measurement time $t$. The remaining inequality is satisfied for $u >1$ when  $c\geq 1$. The condition $u>1$ is automatically fulfilled as we sort the eigenvalues in increasing order, and the condition $c\geq 1$ is fulfilled for all $k\leq i$. Hence, we find that
\begin{align}
	\lambda_k< \mt_i^+, \quad \forall \; \lambda_k \leq \lambda_i.
\end{align} 
Letting $u=\tfrac{\abs{l_{i+1}}^2}{\abs{l_{i}}^2}$ as before, and defining $c'=\tfrac{\abs{l_{i+1}}^2}{\abs{l_{k}}^2}$, we can similarly rewrite the second inequality as, 
\begin{subequations}
\begin{align}
	\label{eq:app-meanvalue-2a}
	&\lambda_k>\mt_{i+1}^+,  \\
	\label{eq:app-meanvalue-2b}
	&\ln{u} > c'\round{1-\frac{1}{u}}  + \frac{\abs{\xt_{\varepsilon}} - \ln{\frac{p_{i+1}}{p_{i}}}}{\gamma t \abs{l_k}^2}. 
\end{align}
\end{subequations}
Again, we may neglect the last term for long measurement times $t$, and we note that the remaining inequality is satisfied for $u>1$ when $c'\leq1$. The latter is fulfilled for all $k\geq i+1$. Accordingly, we find that 
\begin{align}
	\lambda_k>\mt_{i+1}^+, \quad \forall \; \lambda_k \geq \lambda_{i+1}.
\end{align}

A specification of a sufficiently long measurement time in a realistic implementation with a finite error may be obtained by solving the inequalities (\ref{eq:app-meanvalue-1a}) and (\ref{eq:app-meanvalue-2a}) for $t$. This time can be interpreted as a convergence time, which places the mean $\lambda_i$ in the interval $\mt_i^-< \lambda_i < \mt_i^+$ in which $\Pt(i|m)\geq 1-\varepsilon$. The resulting expression for the state $\ket{i}$ is
\begin{align}
\label{eq:app-convergence-time}
	t_i(\varepsilon) \geq \max{ \frac{\abs{\xt_\varepsilon}-\ln{\frac{p_i}{p_{i\pm1}}}}{ \abs{l_{i\pm1}}^2 - \abs{l_i}^2 +  \abs{l_i}^2\ln{\frac{\abs{l_i}^2}{\abs{l_{i\pm1}}^2}}  } }.
\end{align}
This reproduces the results provided in \secref{sec:time-error-relationship}.

\subsection{Vanishing tail probabilities}

The results in the previous section show that all modes in $\Pt(m)$ satisfy one of the conditions for the bounds in \eqref{eq:app-poisson-bounds}. 
We now want to show that these bounds vanish in the long-time limit.

To simplify our notation, we write the bounding jump counts $\mt_i^\pm$ [\eqref{eq:mipm}] as
\begin{align}
    \label{eq:app-mtpm}
    \mt_i^{\pm} = \gamma t \kappa_{i,i\pm1}  - \beta_{i,i\pm1}
\end{align}
where $\kappa_{ij}$ is defined as in \eqref{eq:app-kij}, and we introduced the new parameter
\begin{equation}
	\beta_{ij} = \frac{ \xt_\varepsilon + \ln{ \frac{p_i}{p_j} } }{ \ln{ \frac{\abs{l_i}^2}{\abs{l_j}^2} } }.
\end{equation}

For a general $\mt_i^\pm$, the righthand sides of \eqref{eq:app-poisson-bounds} can be written as
\begin{align*}
    \frac{
    \round{ {\rm e} {\lambda_k}}^{\mt_i^\pm} \exp{-\lambda_k t } }{ \round{{\mt_i^\pm}}^{\mt_i^\pm}} & =\frac{
    \round{ {\rm e}\gamma t \abs{l_k}^2 }^{\mt_i^\pm} \exp{-\gamma t \abs{l_k}^2 + \mt_i^\pm } }{ \round{{\mt_i^\pm}}^{\mt_i^\pm}},\\ 
    & = \curly{\gamma t \abs{l_k}^2  = \frac{ \abs{l_k}^2 }{\kappa_{i,i\pm1}}\round{\mt_i^\pm+\beta_{i,i\pm1}}}, \\
    & \coloneq f(t)h(t),
\end{align*}
where we defined 
\begin{equation}
	f(t) = \round{ \frac{\mt_i^\pm+\beta_{i,i\pm1} }{\mt_i^\pm}}^{\mt_i^\pm}, 
\end{equation}
and 
\begin{equation}
	h(t) = \round{ \tfrac{\abs{l_k}^2}{\kappa_{i,i\pm1}}}^{\mt_i^\pm} \exp{-\gamma t \abs{l_k}^2  + \mt_i^\pm}, 
\end{equation}
at the last equality. 
Next, we want to show that $\displaystyle \lim_{t\to \infty }f(t)$ and $\displaystyle \lim_{t\to \infty }h(t)$ exist. 
Beginning with the limit of $f(t)$, we find that
\begin{align}
    \lim_{t\to \infty} f(t)
    &= \lim_{t\to \infty} \round{ \frac{\mt_i^\pm+\beta_{i,i\pm1} }{ \mt_i^\pm}}^{\mt_i^\pm}, \nonumber\\
    &= \lim_{t \to \infty} \bigg[ \round{ \frac{\gamma t \kappa_{i,i\pm1}}{ \gamma t \kappa_{i,i\pm1} - \beta_{i,i\pm1} }}^{\gamma t \kappa_{i,i\pm1}} \nonumber \\ 
    & \hspace{1.4cm} \round{ \frac{\gamma t \kappa_{i,i\pm1}}{ \gamma t \kappa_{i,i\pm1} - \beta_{i,i\pm1} }}^{-\beta_{i,i\pm1}} \bigg], \nonumber\\
    & = \exp{\beta_{i,i\pm1}} 
    \label{eq:app-ft-limit}
\end{align}
if $\beta_{i,i\pm1}$ is finite. The latter holds for any arbitrarily small but finite error $\varepsilon$, resulting in a correspondingly large but finite $\beta_{i,i\pm1} \propto \mp \abs{\xt_\varepsilon}$.

The limit of $h(t)$ can be evaluated as
\begin{align}
    \lim_{t\to \infty} h(t) & = \lim_{t\to \infty} \round{ \tfrac{\abs{l_k}^2}{\kappa_{i,i\pm1}}}^{\mt_i^\pm} \exp{-\gamma t \abs{l_k}^2  + \mt_i^\pm} \nonumber \\
    & = \lim_{t\to \infty} \exp{-\gamma t \abs{l_k}^2 + \mt_i^\pm \round{1+ \ln{ \frac{\abs{l_k}^2}{\kappa_{i,i\pm1} } } } }, \nonumber\\
    & = \lim_{t\to \infty} \bigg\{ \exp{- \gamma t \round{ \abs{l_k}^2 - \kappa_{i,i\pm1} \round{ 1 + \ln{ \frac{\abs{l_k}^2 }{ \kappa_{i,i\pm1} } } } } } \nonumber \\
    & \hspace{1.4cm} \exp{-\beta_{i,i\pm1} \round{ 1+ \ln{\tfrac{\abs{l_k}^2}{\kappa_{i,i\pm1} } } } } \bigg\}, \nonumber\\
    & = 0, 
    \label{eq:app-ht-limit}
\end{align}
if 
\begin{align}
	\label{eq:app-ht-inequality}
	 \kappa_{i,i\pm1} \round{ \frac{\abs{l_k}^2}{\kappa_{i,i\pm1}}  - 1 - \ln{ \frac{\abs{l_k}^2 }{ \kappa_{i,i\pm1} } } } > 0.
\end{align}
The equality in \eqref{eq:app-ht-inequality} holds if $\tfrac{\abs{l_k}^2}{\kappa_{i,i\pm1}} \neq 1$. For the first bound for the modes satisfying $\lambda_k\leq\lambda_i< \mt_i^+$, this condition translates to
\begin{align} \label{eq:app-vanishing-condition1}
	&\abs{l_k}^2 \neq \kappa_{i,i+1}, \nonumber \\
	&\ln{\frac{\abs{l_{i}}^2}{\abs{l_{i+1}}^2}} \neq \frac{\abs{l_i}^2}{\abs{l_k}^2} \round{1-\frac{\abs{l_{i+1}}^2}{\abs{l_i}^2}} ,
\end{align}
which is always fulfilled as, in this case, $\tfrac{\abs{l_{i}}^2}{\abs{l_{i+1}}^2}< 1$ and $\frac{\abs{l_i}^2}{\abs{l_k}^2}\geq 1$, c.f. $\ln{u} \neq c (1- u^{-1})$ with $c \geq 1$ that does not have any solutions when $u< 1$. The second bound instead concerns the modes with $\lambda_k \geq \lambda_{i+1} > \mt_{i+1}^-$. In this case, the condition is
\begin{align}\label{eq:app-vanishing-condition2}
	&\abs{l_k}^2 \neq \kappa_{i+1,i} .
\end{align}
Since $\kappa_{i+1,i} = \kappa_{i,i+1}$, this leads to a similar condition of the form $\ln{u} \neq c (1- u^{-1})$, but this time with $c < 1$, as $\abs{l_k}^2>\abs{l_i}^2$ in this case. This case likewise lacks solutions for $u<1$. Hence, we find that the condition in \eqref{eq:app-ht-inequality} is satisfied for the bounds in \eqref{eq:app-poisson-bounds} for all intervals $i \in (1,d-1)$.
Finally, employing the results in Eqs.~(\ref{eq:app-ht-limit}) and (\ref{eq:app-ft-limit}), we find that 
\begin{equation}
    \lim_{t\to \infty}\frac{
    \round{ {\rm e} {\lambda_k}}^{\mt_i^+} \exp{-\lambda_k t }}{\round{\mt_i^+}^{\mt_i^+}} =  \lim_{t\to \infty}\frac{
    \round{ {\rm e} {\lambda_k}}^{\mt_{i+1}^-} \exp{-\lambda_k t }}{\round{\mt_{i+1}^-}^{\mt_{i+1}^-}} = 0
\end{equation}
for all $i \in (1,d-1)$, which was the goal of this section.

\section{Convergence rate for diffusive trajectories}\label{app:Kijk-diffusive}
Here we provide the detailed calculations that derive the rate in \eqref{eq:Ri-diffusive} in the main text. Making the replacements $l_i \to l_i + \alpha$ in \eqref{eq:convergence-rate}, we obtain the expression
\begin{equation}
    \label{eq:Kijk-alpha-appendix}
    R_{ii\pm1}^\alpha \hspace{-0.05cm}= \abs{l_{i\pm1}\hspace{-0.05cm}+\hspace{-0.03cm}\alpha}^2\hspace{-0.1cm} - \abs{l_i\hspace{-0.03cm}+\hspace{-0.03cm}\alpha}^2 \hspace{-0.1cm} + \abs{l_i\hspace{-0.03cm}+\hspace{-0.03cm}\alpha}^2 \ln{\hspace{-0.05cm}\frac{\abs{l_{i}\hspace{-0.03cm}+\hspace{-0.03cm}\alpha}^2}{\abs{l_{i\pm1}\hspace{-0.05cm}+\hspace{-0.03cm}\alpha}^2}\hspace{-0.05cm}}
\end{equation}
for the rates of distinguishing the $ith$ mode from the two neighbouring modes $i\pm1$.
We consider the general case where $\alpha \in \mathbb{C}$ and $l_i \in \mathbb{C}$ and will use the notation
\begin{align}
	 l_i
	  =x_i+i y_i ,
\end{align}
and
\begin{subequations}
\begin{align}
	 \alpha 
	  &=\alpha_x+i \alpha_y, \\
	 &=\abs{\alpha}\exp{i \phi_\alpha} , \\
	  &=\abs{\alpha}\cos( \phi_\alpha)+i\abs{\alpha}\sin( \phi_\alpha), 
\end{align}
\end{subequations}
where the different expressions for $\alpha$ will be employed as most convenient. The first two terms in \eqref{eq:Kijk-alpha-appendix} can be expanded as
\begin{widetext}
\begin{equation}
\label{eq:first-two-terms}
	\abs{l_{i\pm1}+\alpha}^2 - \abs{l_i+\alpha}^2 = \abs{l_{i\pm1}}^2 - \abs{l_i}^2
	+2\abs{\alpha}\bigg((x_{i\pm1}-x_i)\cos{\phi_\alpha}+(y_{i\pm1}-y_i)\sin{\phi_\alpha}\bigg).
\end{equation}
Next, we note that
\begin{subequations}
\begin{align}
	\ln{\abs{l_i+\alpha}^2} &= 2 \ln{\abs{(x_i+\alpha_x)+i(y_i+\alpha_y)}} \\
	&= 2 \ln{\sqrt{(x_i+\alpha_x)^2+i(y_i+\alpha_y)^2}} \\
	&= \ln{\hard{ \abs{l_i}^2+2(x_i\alpha_x+y_i\alpha_y)+\abs{\alpha}^2}} \\
	&= \ln{\abs{\alpha}^2}+\ln{ \hard{1+\frac{2}{\abs{\alpha}}\round{ x_i\cos{\phi_\alpha} + y_i\sin{\phi_\alpha}} + \frac{\abs{l_i}^2}{\abs{\alpha}^2} }}, 
\end{align}
\end{subequations}
where the second logarithm in the last expression is of the form $\ln{[1+u]}$ with $\abs{u}<1$ when $\abs{\alpha}\gg1$. Hence, we may expand it as $\ln{[1+u]}=u-\tfrac{u^2}{2}+\mathcal{O}(u^3)$. Evaluating the cubic term $u^3$, we find that $\mathcal{O}(u^3)=\mathcal{O}(\frac{1}{\abs{\alpha}^3})$. Therefore, we may write the logarithm of $\abs{l_i+\alpha}^2$ as
\begin{equation}
	\ln{\abs{l_i+\alpha}^2} = \ln{\abs{\alpha}^2} + \frac{2}{\abs{\alpha}} \round{ x_i\cos{\phi_\alpha} + y_i\sin{\phi_\alpha}} + \frac{\abs{l_i}^2}{\abs{\alpha}^2} - \frac{2}{\abs{\alpha}^2} \round{ x_i\cos{\phi_\alpha} + y_i\sin{\phi_\alpha}}^2 +\mathcal{O}\round{\frac{1}{\abs{\alpha}^3}}.
\end{equation}
With this expansion, we may now write the third term in \eqref{eq:Kijk-alpha-appendix} as
\begin{subequations}
\begin{align}
	\abs{l_i+\alpha}^2\ln{\frac{\abs{l_i+\alpha}^2}{\abs{l_{i\pm1}+\alpha}^2}} 
	=& \round{\abs{l_i}^2+2\abs{\alpha} \round{ x_i\cos{\phi_\alpha} + y_i\sin{\phi_\alpha} } +\abs{\alpha}^2} \round{\ln{\abs{l_{i}+\alpha}^2} -\ln{\abs{l_{i\pm1}+\alpha}^2} } \\
	=& \hspace{0.05cm} 2\abs{\alpha}^2 \bigg( \round{x_i-x_{1\pm1}}\cos{\phi_\alpha} + \round{y_i-y_{i\pm1}}\sin{\phi_\alpha} \bigg) +\abs{l_i}^2-\abs{l_{i\pm1}}^2 \nonumber \\
	& -2\big( \round{ x_i\cos{\phi_\alpha} + y_i\sin{\phi_\alpha}}^2 -\round{ x_{i\pm1}\cos{\phi_\alpha} + y_{i\pm1}\sin{\phi_\alpha}}^2 \big) \nonumber \\
	&+ 4\round{ x_i\cos{\phi_\alpha} + y_i\sin{\phi_\alpha} } \big( \round{x_i-x_{i\pm1}}\cos{\phi_\alpha} + \round{y_i-y_{i\pm1}}\sin{\phi_\alpha} \big) \nonumber \\
	&
	+D_{ \abs{\alpha},\abs{\alpha}^2 } + \mathcal{O}\round{\frac{1}{\abs{\alpha}^3}}.
	\label{eq:third-term}
\end{align}
\end{subequations}
Here, we have grouped all terms proportional to $\tfrac{1}{\abs{\alpha}}$ and $\tfrac{1}{\abs{\alpha}^2}$ into the parameter $D_{ \abs{\alpha},\abs{\alpha}^2 }$. Then, by adding the results in \eqref{eq:third-term} \cg{to} \eqref{eq:first-two-terms} and taking the limit $\abs{\alpha}\to \infty$, we obtain the expression 
\begin{align}
    \lim_{\abs{\alpha}\to \infty} R_{ii\pm1}^{\alpha} =& \hspace{0.06cm} 2 \bigg| \hard{ \round{ x_{i\pm1} \cos{\phi_\alpha}  +  y_{i\pm1}\sin{\phi_\alpha} }^2  - \round{ x_i\cos{\phi_\alpha}+y_i \sin{\phi_\alpha} }^2 } \nonumber \\  
     & + 4\hard{x_i\cos(\phi_\alpha) + y_i\sin(\phi_\alpha)} 
     \hard{ \round{x_i-x_{i\pm1}}\cos(\phi_\alpha) +  \round{y_i-y_{i\pm1}} \sin(\phi_\alpha)}\bigg|,
\end{align}
which gives the expression in \eqref{eq:Ri-diffusive} in the main text. 
\end{widetext}

\section{Asymptotic convergence in the large-\texorpdfstring{$m$}{m} limit}\label{sec:app-fixed-m-perspective}
In this appendix, we show that all trajectories generated by the measurement (\eqref{eq:weak-measurement}) converge to a single eigenstate in the large-$m$ limit when conditioning on the jump count $m$, rather than on the externally controlled measurement time. To do this, we make the same assumptions as in the main text. That is, for a $d$-dimensional Hilbert space, we assume that the jump operator $\L$ is diagonalisable in an orthonormal basis of eigenstates $\curly{\ket{i}}_{i=1}^d$ that coincide with the eigenbasis of the Hamiltonian, i.e $[\H,\L]=0$.  To improve the reading experience, we restate the relevant equations from the main text. 

The first object we need for our proof is the first-passage-time distribution 
\begin{subequations}
	\begin{align}
		\Pm(t) 	&= \sum_{i=1}^d p_i \frac{ t^{m-1} \round{\gamma \abs{l_i}^2 }^m }{(m-1)!} \exp{-\gamma t \abs{l_i}^2} \\
				&= \sum_{i=1}^d p_i \Erl(m,\Gamma_i;t),
	\end{align}
\end{subequations}
provided in \eqref{eq:Pmt} of the main text.
The expression on the second line highlights that it can be written as a weighted sum of Erlang distributions, analogous to the Poissonian modes in \eqref{eq:Ptm}.  The Erlang distribution is characterised both by its shape parameter, which here is the jump count $m$, and the rate $\Gamma_i=\gamma \abs{l_i}^2$, while the corresponding mean of the $i$th mode is given by $\mu_i=m/\Gamma_i$.

Next, as noted in \eqref{eq:Pcond-m} in the main text, the conditional probability distribution $\Pm(i|t)$ in the fixed-$m$ protocol has the same form as $\Pt(i|m)$ in the fixed-$t$ protocol. That is, 
\begin{align}
	\Pm(i | t) =  \frac{ 1 }{ 1 + \displaystyle \sum_{j \neq i} \exp{x_{ij}} },
\end{align}
where the generalised variables are defined in \eqref{eq:xij} as
\begin{equation}
        x_{ij}= \gamma t \round{\abs{l_i}^2\hspace{-1pt} -\abs{l_j}^2}\hspace{-0.5pt} - m \ln{\frac{\abs{l_i}^2}{\abs{l_j}^2}} \hspace{-0.5pt} - \ln{\frac{p_i}{p_j}}. 
\end{equation}
It follows that the same non-degeneracy condition (\eqref{eq:non-degeneracy-condition}) is necessary to enable asymptotic projection onto a single eigenstate. Thus, we assume in the following that all eigenvalues of $\L^\dagger\L$ are distinct and we label them such that $\abs{l_1}^2<\abs{l_2}^2< \cdots < \abs{l_d}^2$.

The convergence condition is straightforwardly stated as
\begin{equation}
    \lim_{m\to \infty} \frac{1}{d-1}\sum_{i=1}^{d-1}\sum_{j=i+1}^d \abs{\Pm( i |t)-\Pm(j | t)}= 1, 
\end{equation}
where
\begin{equation}
    \abs{\Pm\round{ i |t}-\Pm\round{ j |t}} = 
        \begin{cases}
            1 & \text{if} \; \ket{\psi}_t=\ket{i \,{\rm or}\,j}, \\
            0 & \text{if} \; \ket{\psi}_t=\ket{k \neq i,j}, \\
            \in\mleft(0,1\mright) & \text{otherwise},
        \end{cases}
\end{equation}
analogously to Eqs. (\ref{eq:convergence}) and (\ref{eq:pairwise-difference}).
As before, we also note that 
\begin{align}
	\Pm(i | t) \geq \frac{1}{1+(d-1)\exp{\xt_i} },
\end{align}
with $\xt_i = \max{x_{ij_1}, x_{ij_2}, ..., x_{ij_{d-1}}}$, and we solve
\begin{align}
	\Pm(i | t) \geq 1-\varepsilon.
\end{align}
while substituting $\xt_i = \xt_\varepsilon$ to obtain the threshold 
\begin{equation}
	\xt_\varepsilon = \ln{\frac{\varepsilon}{(d-1)(1-\varepsilon)}}. 
\end{equation}

Now, we want to determine $\xt_i$ for the values of $m$ where $\Pm(i|t)\approx 1$. Since we later want to take the limit $m\to \infty$, we characterise $\xt_i$ as a function of the normalised variable $\tfrac{\gamma t}{m}$, which is reciprocal to $\tfrac{m}{\gamma t}$, characterising the functional behaviour of $\xt_i$ in the fixed-$t$ protocol.  For large $m$, one can easily show that 
(c.f. \eqref{eq:xmax})
\begin{align}
	\xt_i = 
	\begin{cases}
		&x_{i,i+1}, \quad  \text{for}\; \; \frac{1}{\kappa_{i+1,i+2}} <\frac{\gamma t}{m} < \frac{1}{\kappa_{i-1,i+1}} , \\
		&x_{i,i-1}, \quad \text{for}\; \; \frac{1}{\kappa_{i-1,i+1}}  < \frac{\gamma t}{m} < \frac{1}{\kappa_{i-2,i-1}}. 
	\end{cases}
\end{align}
These relationships follow directly from Eqs.~(\ref{eq:app-prop1})-(\ref{eq:app-prop5}) with the only difference that we assume $m \gg \ln{\frac{p_{j'}}{p_j}}/\round{\abs{l_{j'}}^2-\abs{l_{j}}^2}$ instead of assuming large $\gamma t$ to obtain the analogue of \eqref{eq:app-prop3}. 

Continuing to follow our previous steps, we then solve $x_{i,i\pm1}=\xt_\varepsilon$ for $t$ to obtain the threshold first-passage times
\begin{equation}
	\gamma \tilde{t}_i^\pm = \frac{m \ln{\frac{\abs{l_i}^2}{\abs{l_{i\pm1}}^2}} + \xt_\varepsilon + \ln{\frac{p_i}{p_{i\pm1}}} }{  \abs{l_i}^2 - \abs{l_{i\pm1}}^2},
\end{equation}
which defines the interval $\tilde{t}_i^+ \leq t \leq \tilde{t}_i^-$ within which the bound $\Pm(i | t) \geq 1-\varepsilon$ is satisfied. 
With this characterisation, we may proceed and evaluate the cumulative probability of finding $t$ outside these intervals. The inequality analogue to \eqref{eq:cumulative-bound} is 
\begin{align} 
		\Pm(\tilde{t}_{i+1}^-\hspace{-0.07cm} \leq \hspace{-0.03cm} t \hspace{-0.03cm}\leq \hspace{-0.03cm} \tilde{t}_{i}^+ )
		&\leq 
         \hspace{-0.3cm}\sum_{k: \mu_k < \tilde{t}_{i+1}^-} \hspace{-0.3cm} p_k \Erl(m, \Gamma_k; t \geq \tilde{t}_{i+1}^- ) \nonumber \\
		& \; + \hspace{-0.3cm} \sum_{k: \mu_k > \tilde{t}_{i}^+}  \hspace{-0.25cm} p_k \Erl(m, \Gamma_k; t \leq \tilde{t}_{i}^+ ), 
\end{align}
and the two cumulative tail probabilities of the Erlang distributions are bounded by the Chernoff bounds~\cite{mitzenmacher_probability_2005_ch2} according to
\begin{subequations}\label{eq:app-chernoff-erlang}
\begin{align}
\label{eq:app-chernoff-erlang1}
    \Erl(m, \Gamma_k; t \geq \tilde{t}_{i+1}^- ) 
    \hspace{-0.02cm}\leq \hspace{-0.035cm}\round{\hspace{-0.02cm}\frac{\Gamma_k \tilde{t}_{i+1}^-}{m}\hspace{-0.03cm}}^m \hspace{-0.1cm}\exp{-(\Gamma_k \tilde{t}_{i+1}^- - m)},
\end{align}
for $\mu_k < \tilde{t}_{i+1}^-$, and
\begin{align}
\label{eq:app-chernoff-erlang2}
    \Erl(m, \Gamma_k; t \leq \tilde{t}_{i}^+ ) 
    \hspace{-0.02cm}\leq \hspace{-0.035cm}\round{\hspace{-0.02cm}\frac{\Gamma_k \tilde{t}_{i}^+}{m}\hspace{-0.03cm}}^m \hspace{-0.1cm}\exp{-(\Gamma_k \tilde{t}_{i}^+ - m)},
\end{align}
\end{subequations}
for $\mu_k > \tilde{t}_{i}^+$. 
Beginning with the first inequality $\mu_k < \tilde{t}_{i+1}^-$, we find that it is satisfied $\forall k\geq i+1$. To see this, recall that $\mu_k=\tfrac{m}{\Gamma_k}=\tfrac{m}{\gamma \abs{l_k}^2}$ and assume $m\gg \abs{\xt_\varepsilon}-\ln{\tfrac{p_i}{p_{i+1}}}$. Then, we may rewrite the inequality on the form $\ln{u}>c(1-\tfrac{1}{u})$ with $u\equiv \tfrac{\abs{l_{i+1}}^2}{\abs{l_{i}}^2}>1$ and $c\equiv \tfrac{\abs{l_{i+1}}^2}{\abs{l_{k}}^2}$ and it is easily confirmed that this inequality holds $\forall c \leq 1$, which is satisfied $\forall k \geq i+1$. Similarly, the inequality $\mu_k > \tilde{t}_{i}^+$ can be rewritten as $\ln{\tfrac{1}{u}}>c(1-u)$ with $u\equiv \tfrac{\abs{l_{i+1}}^2}{\abs{l_{i}}^2}>1$ and $c\equiv \tfrac{\abs{l_{i}}^2}{\abs{l_{k}}^2}$, again assuming that $m\gg \abs{\xt_\varepsilon}-\ln{\tfrac{p_i}{p_{i+1}}}$. This inequality holds $\forall c \geq 1$, which is satisfied $\forall k \leq i$. 

It now remains to show that the right-hand sides of Eqs.~(\ref{eq:app-chernoff-erlang1}) and (\ref{eq:app-chernoff-erlang2}) vanish in the limit $m\to \infty$.
To demonstrate this, we rewrite the threshold times as
\begin{equation}
    \tilde{t}_{i}^{\pm} = m \frac{1}{\gamma \kappa_{i,i\pm1}}  + \zeta_{i,i\pm1},
\end{equation}
where we have defined
\begin{equation}
    \zeta_{ij} = \frac{\xt_\varepsilon+\ln{\frac{p_i}{p_j}}}{\gamma(\abs{l_i}^2-\abs{l_j}^2)}.
\end{equation}
For a general $\tilde{t}_i^\pm$, the right-hand sides of \eqref{eq:app-chernoff-erlang} can be rewritten as
\begin{align}
    \round{\frac{\Gamma_k \tilde{t}_{i}^\pm}{m}}^m
     \exp{-(\Gamma_k \tilde{t}_{i}^\pm - m)}  
     =f(m) h(m),
\end{align}
where
\begin{align} \label{eg:app-fm-limit}
    \lim_{m\to\infty} f(m) &= \lim_{m\to\infty}      \round{\frac{m + \gamma \kappa_{i,i\pm1}\zeta_{i,i\pm1}}{m}}^m  \nonumber \\
    &= \exp{\gamma \kappa_{i,i\pm1}\zeta_{i,i\pm1}},
\end{align}
and
\begin{align}
    \lim_{m\to\infty} h(m) &= \exp{-m(\tfrac{\Gamma_k}{\gamma \kappa_{i,i\pm1}}-1-\ln{\tfrac{\Gamma_k}{\gamma \kappa_{i,i\pm1}}})}
    \exp{-\Gamma_k\zeta_{i,i\pm1}} \nonumber \\
    &= 0,
\end{align}
if $\tfrac{\Gamma_k}{\gamma \kappa_{i,i\pm1}}-1-\ln{\tfrac{\Gamma_k}{\gamma \kappa_{i,i\pm1}}}> 0$. The latter inequality holds for all $\tfrac{\Gamma_k}{\gamma \kappa_{i,i\pm1}} \neq 1$. Recalling that $\Gamma_k=\gamma \abs{l_k}^2$, and specifying the threshold times $\tilde{t}_{i+1}^-$ and $\tilde{t}_{i}^+$ we find that this condition is the same as in the previous proof and was demonstrated to hold in the text surrounding Eqs.~(\ref{eq:app-vanishing-condition1}) and (\ref{eq:app-vanishing-condition2}). 

Since $\zeta_{i,j}$ is finite for all $i$ and $j$ when the error $\varepsilon$ is small but finite, it follows that \eqref{eq:app-ft-limit} and $\exp{-\Gamma_k\zeta_{i,i\pm1}}$ are also finite . Thus, the right-hand sides of \eqref{eq:app-chernoff-erlang} are 0 in the limit $m\to\infty$. 
Finally, these results allow us to conclude that 
\begin{align}
    \lim_{\varepsilon \to 0} \lim_{m\to \infty} \sum_{i=1}^{d-1} \Pm(\tilde{t}_{i+1}^- \leq  t \leq \tilde{t}_{i}^+ ) = 0,
\end{align}
demonstrating that all trajectories generated by the measurement (\eqref{eq:weak-measurement}) converge to a single eigenstate in the large-$m$ limit.

\FloatBarrier
\bibliography{references}

\end{document}